\documentclass[pdftex,twocolumn,epjc3]{svjour3}          % twocolumn

\RequirePackage[T1]{fontenc}

\smartqed  % flush right qed marks, e.g. at end of proof

\RequirePackage{graphicx}
\RequirePackage{mathptmx}      % use Times fonts if available on your TeX system
\RequirePackage{flushend}
\RequirePackage[numbers,sort&compress]{natbib}
\RequirePackage[colorlinks,citecolor=blue,urlcolor=blue,linkcolor=blue]{hyperref}
\usepackage{amsmath}
\usepackage{siunitx}
\usepackage[version=4]{mhchem}
\usepackage{graphicx}  % for including graphics
\usepackage{subcaption}  % for subfigures
\usepackage{tikz}
\usepackage{xspace}
\DeclareSIUnit\bar{bar}
\DeclareSIUnit\voltspercm{V/cm}
\DeclareSIUnit\kilovoltspercm{kV/cm}
\DeclareSIUnit\mbarliterpersec{mbar l/s}

\newcommand{\kr}{$^{\text{83m}}$Kr\xspace}

\journalname{Eur. Phys. J. C}

\begin{document}

\title{Operating a large-diameter dual-phase liquid xenon TPC \\ in the unshielded PANCAKE facility}

\author{
Julia M\"uller\thanksref{e1} \and 
Jaron Grigat\thanksref{e2} \and  
Robin Glade-Beucke \and 
Sebastian Lindemann \and
Tiffany Luce \and
Gnanesh Chandra Madduri \and
Jens Reininghaus \and
Marc Schumann \and
Adam Softley-Brown\thanksref{aa1} \and
Andrew Stevens
}

\thankstext{e1}{e-mail: julia.mueller@physik.uni-freiburg.de}
\thankstext{e2}{e-mail: jaron.grigat@physik.uni-freiburg.de}
\thankstext{aa1}{now at: University of Sheffield, UK}
%\thankstext{aa2}{also at: SVNIT, Surat, India}

\institute{Physikalisches Institut, Albert-Ludwigs-Universität Freiburg, Hermann-Herder-Straße 3, 79104 Freiburg,
Germany
}

\date{Received: date / Accepted: date}

\maketitle

\sloppy

\begin{abstract}
Future liquid-xenon (LXe) based observatories for rare processes, such as XLZD, require testing of large components and sub-assemblies in cryogenic liquid or gaseous xenon environments. Here we present results from the stable operation of a shallow dual-phase LXe TPC with an inner diameter of 133.4\,cm and a height of 3.1\,cm in the unshielded PANCAKE platform, without underground suppression of cosmic-ray backgrounds. A total of 340\,kg of xenon was used in the experiment, of which 127\,kg constituted the active TPC mass. Measurements of the LXe purity-dependent electron lifetime and the electron drift velocity in LXe demonstrate that sensitive measurements to characterize the TPC performance are possible in a high-background environment, even with a very basic PMT-based light detection system. Improving this will straightforwardly reduce the TPC threshold, which was observed to be around 15\,keV for electronic recoils in TPC operation.

\end{abstract}

%%%%%%%%%%%%%%%%%%%%%%%%%%
% main text starts here
%      -> please only edit the file pancake.tex!!

%%%%%%%%%%%%%%%%%%%%%%%%%%%%%%%%%%%%%%%%%%%%%%%%%%%%%%%%%%
%%%%%%%%%%%%%%%%%%%%%%%%%%%%%%%%%%%%%%%%%%%%%%%%%%%%%%%%%%
%%%%%%%%%%%%%%%%%%%%%%%%%%%%%%%%%%%%%%%%%%%%%%%%%%%%%%%%%%
\section{Introduction} 

Dual-phase liquid xenon time projection chambers (LXe TPCs) have led the direct search for dark matter in the form of WIMPs with masses $\geq$5\,GeV/$c^2$ for more than a decade~\cite{Billard:2021uyg}, and several detectors at the multi-ton scale are currently in operation~\cite{XENON:2024wpa,LZ:2019sgr,PandaX:2024qfu}. In order to explore the entire accessible parameter space before the sensitivity is seriously impacted by irreducible neutrino interactions (the so-called neutrino floor~\cite{Billard:2013qya} or fog~\cite{OHare:2021utq}), exposures of at least 200\,t\,$\times$\,y are required~\cite{Schumann:2015cpa}. Two detector concepts capable of delivering such exposures are currently being discussed: PandaX-xt~\cite{PANDA-X:2024dlo} and XLZD~\cite{XLZD:2024nsu}, where XLZD plans to collect an exposure of up to 1000\,t\,$\times$\,y in case a hint of a signal is seen at smaller exposures. Due to its large target and very low background level, an instrument such as XLZD offers excellent sensitivity to many other science channels~\cite{Aalbers:2022dzr}, in particular neutrinoless double-beta decay of $^{136}$Xe~\cite{XLZD:2024pdv} and precision measurements of low-energy solar neutrinos~\cite{DARWIN:2020bnc}.

The cylindrical XLZD TPC will have linear dimensions of about 3\,m, which is about a factor of two larger than the current generation of detectors. Large-scale test facilities are therefore required to develop and test critical components for XLZD at full scale and under realistic cryogenic conditions. The PANCAKE test platform is one such facility and allows for tests of large, flat TPC components~\cite{Brown:2023vgf}. It is installed in an unshielded environment at the University of Freiburg, Germany, and was not constructed from dedicated low-background materials. As a result, it does not provide the low-background conditions required for rare event searches. In this work, we present the first operation of a shallow dual-phase LXe TPC with a diameter of about 1.33\,m in PANCAKE, demonstrating the diagnostic capabilities of the platform. The PANCAKE platform and its instrumentation are presented in Sect.~\ref{sec::setup}, and details on the installed TPC are given in Sect.~\ref{sec::TPC}. Additional instrumentation needed for the TPC operation is presented in Sect.~\ref{sec::tools}. The operation of the TPC is described in Sect.~\ref{sec::operation}, and some key results illustrating the capabilities of the facility are presented in Sect.~\ref{sec::results}. 

%%%%%%%%%%%%%%%%%%%%%%%%%%%%%%%%%%%%%%%%%%%%%%%%%%%%%%%%%%
%%%%%%%%%%%%%%%%%%%%%%%%%%%%%%%%%%%%%%%%%%%%%%%%%%%%%%%%%%
%%%%%%%%%%%%%%%%%%%%%%%%%%%%%%%%%%%%%%%%%%%%%%%%%%%%%%%%%%
\section{Experimental Setup}
\label{sec::setup}

The PANCAKE facility is the world’s largest liquid xenon test platform to date, featuring a flat floor with an internal diameter of \qty{2.75}{m}.
A detailed description of the facility, which currently has a xenon inventory of \qty{400}{kg}, and of its commissioning can be found in~\cite{Brown:2023vgf}. Here, we highlight the components relevant to the TPC operation and characterization in this work.

\begin{figure}
    \centering
    \includegraphics[width=\linewidth]{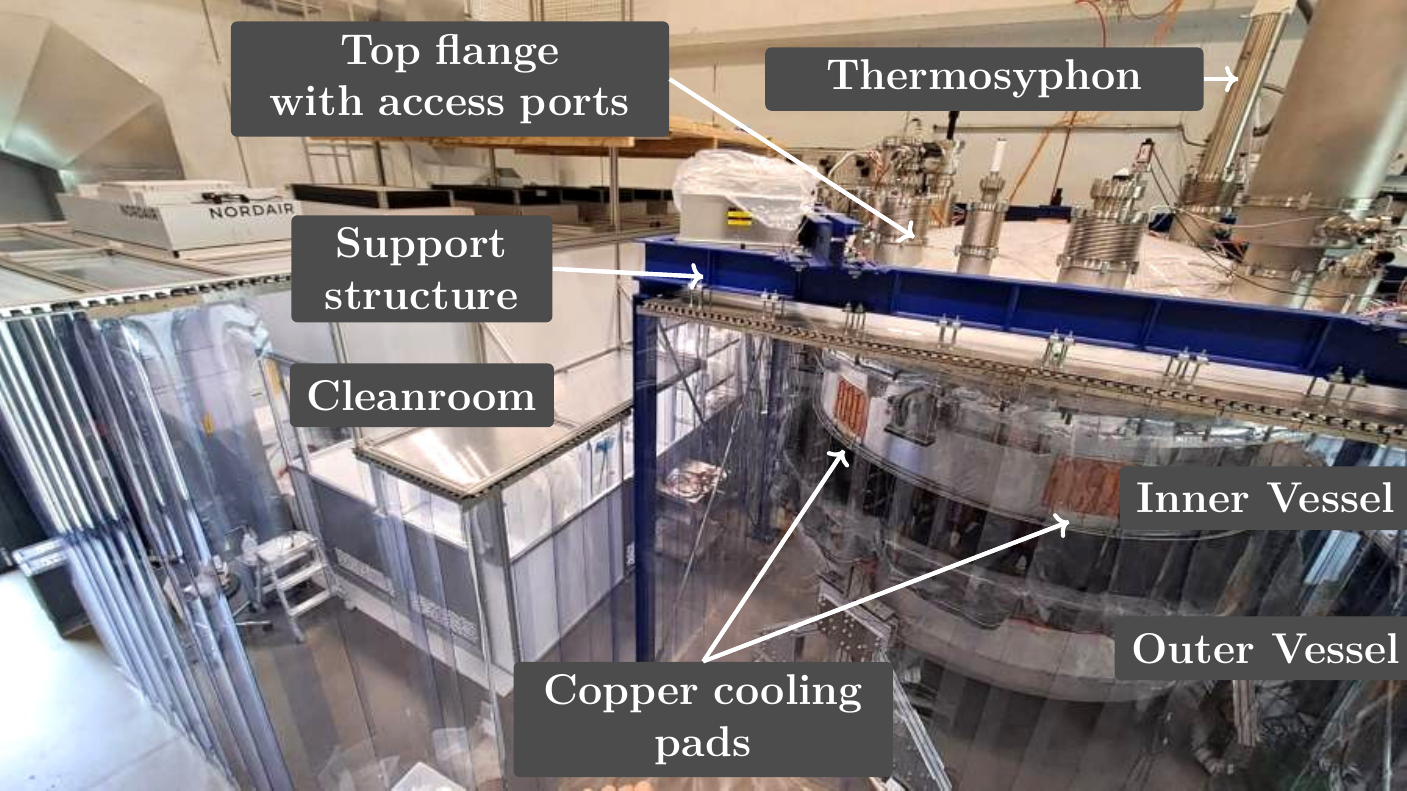}
    \caption{The PANCAKE facility on the right with its surrounding soft-wall clean area and adjacent cleanroom. The outer cryostat vessel was lowered and the (closed) inner vessel is visible. The cryostat is suspended from the blue support structure. }
    \label{fig:overview_pancake_with_cleanrooms}
\end{figure}

PANCAKE houses a vacuum-insulated stainless steel cryostat consisting of an inner and an outer vessel. Figure~\ref{fig:overview_pancake_with_cleanrooms} shows a photograph of the facility at the University of Freiburg. The photograph was taken while the bottom part of the outer vessel was lowered to allow a view of the inner vessel. The top lids of the cryostat vessels are suspended from a support structure on three M48 rods, which allow the detector to be leveled.

The inner vessel is designed to hold liquid xenon at gas pressures of up to \qty{3}{bar} absolute. It was sealed with a \SI{2}{mm} oxygen-free high-conductivity copper wire compressed between the cryostat flanges, yielding acceptable leak rates~\cite{Brown:2023vgf}. Its flat floor reduces the amount of liquid xenon required to submerge large-diameter components during testing. It was demonstrated in Ref.~\cite{Brown:2023vgf} that the effective cryostat diameter used for LXe tests can be reduced by installing a cylindrical open-topped stainless steel vessel with a flat floor in the inner cryostat. Thermally insulating this vessel using poly-tetrafluoroethylene (PTFE) standoffs allows collecting LXe only in the open-topped vessel and not on the inner cryostat's floor. 

On top of the cryostat, two CF-150 ports and six CF-63 ports provide access to the inner vessel. These ports pass through the insulation vacuum and are equipped with feedthroughs to connect instrumentation and accessories, with gas pipes and a pumping port. Four ports connect to the outer vessel. The inner and outer vessels are evacuated using two independent Leybold Turbovac 450i turbomolecular pumps. An MDC AV-600M CF-150 angle valve installed between the turbomolecular pump and the inner vessel allows the evacuation port to be closed when the vessel is filled with xenon gas.

Thermal insulation of the inner vessel from the environment is achieved using support fixtures designed for minimal thermal conduction, by evacuating the volume between the inner and outer vessels, and by installing RUAG Space COOLCAT~2 NW superinsulation, which consists of 30 layers of \qty{12}{\um} thick polyester foil.
This effectively minimizes the radiative heat load, resulting in a total cooling power requirement of approximately \qty{100}{W} when the vessel is filled with liquid xenon.

Two separate liquid nitrogen-based cooling systems are available. The first is used to cool the steel of the inner cryostat from room temperature to approximately LXe temperatures via copper cooling pads directly attached to the stiffening structure of the lower inner vessel. 
The large heat capacity of the vessel dampens the effect of temperature variations in the cooling system. Two of the pads can be seen in Fig.~\ref{fig:overview_pancake_with_cleanrooms}. The system provides a cooling power of up to \qty{600}{W}. This is more than sufficient to meet the \qty{100}{W} requirement. For stable detector operation, only the second system, based on a thermosyphon, is used. It continuously liquefies xenon gas with an adjustable cooling power of typically \qty{150}{W}.

Approximately \qty{400}{kg} of xenon are currently stored in a gas storage system consisting of six 50-liter gas bottles that can be cooled to recover xenon from the cryostat via cryopumping, as well as six bottles used to store warm gas. The gas of any bottle can be fed to the inner cryostat via the central gas panel. 
To purify the xenon gas, the gas panel provides a KNF~N143~SP.12E diaphragm pump, an SAES Mono Torr PS4-MT15-R rare gas purifier, and a Teledyne Hastings HFC-D-303-B flow controller. 
These components are used during operation to purify the xenon gas. To reduce the heat load, the purification loop contains a custom-made heat exchanger. 
The gas panel also allows for the installation of temporary equipment through a series of valves and bypasses. 
For this work, a \kr conversion electron source~\cite{Kastens:2009pa,Hannen:2011mr} and a Tiger Optics HALO+ trace gas analyzer to measure H$_{2}$O in xenon gas at parts-per-billion (ppb) levels were installed. 

In order to provide a dust-free environment for sensitive high-voltage tests and to avoid contamination of detector components, 
a commercial hard-wall ISO class~6 clean room with a footprint of \qty{10}{\meter\squared} was installed next to the cryostat as can be seen in Fig.~\ref{fig:overview_pancake_with_cleanrooms}. 
A cleanroom extension made of polycarbonate panels on top and soft-wall curtains on the sides surrounds the support structure and extends to the ISO-6 clean room. To further reduce the dust ingress from outside the cleanroom between the curtains, thin polyethylene (PE) foil was attached to the areas that did not need to be opened. Three HEPA fan filter units were also installed on the extension to provide airflow.

The PANCAKE facility uses the open-source slow control framework \emph{Doberman}~\cite{Zappa:2016zsn} to read out all auxiliary sensors installed across the facility. Doberman has been developed over the past decade and was specifically adapted to the requirements of PANCAKE: it interfaces with 34~devices and continuously monitors more than 300~parameters. An effective alarm distribution system, remote control capabilities, and automated feedback pipelines enable stable operation of the experiment by a relatively small team.

%%%%%%%%%%%%%%%%%%%%%%%%%%%%%%%%%%%%%%%%%%%%%%%%%%%%%%%%%%
\section{Time Projection Chamber}
\label{sec::TPC}

The TPC operated in this work had an cylindrical active volume with a diameter of 1334\,mm and a height of 31\,mm, enclosing  about 127\,kg of LXe. It was built from three stainless steel electrodes, cathode, gate and anode, with support frames of $\sim$\SI{1.4}{\meter} outer diameter. These three electrodes were built and tested to upgrade the XENONnT dark matter experiment~\cite{Elykov:2025nri}. They were biased independently and no field-shaping electrodes were installed. The anode and gate were parallel wire electrodes with 265~and 266~stainless steel wires of \SI{216}{\um} diameter (California Fine Wire), stretched along 24-sided polygonal stainless steel frames with thicknesses of \SI{24}{\mm} and \SI{20}{\mm}, respectively. The wire spacing was 5\,mm. 
The cathode consisted of an etched hexagonal stainless steel mesh with a thickness of \SI{300}{\um} and cell width of \SI{7.5}{\mm}, clamped to a circular stainless steel frame of \SI{20}{\mm} thickness. The clamps increase the overall thickness to \SI{25}{\mm}. 

\begin{figure}[!h]
    \centering
    \includegraphics[width=\linewidth]{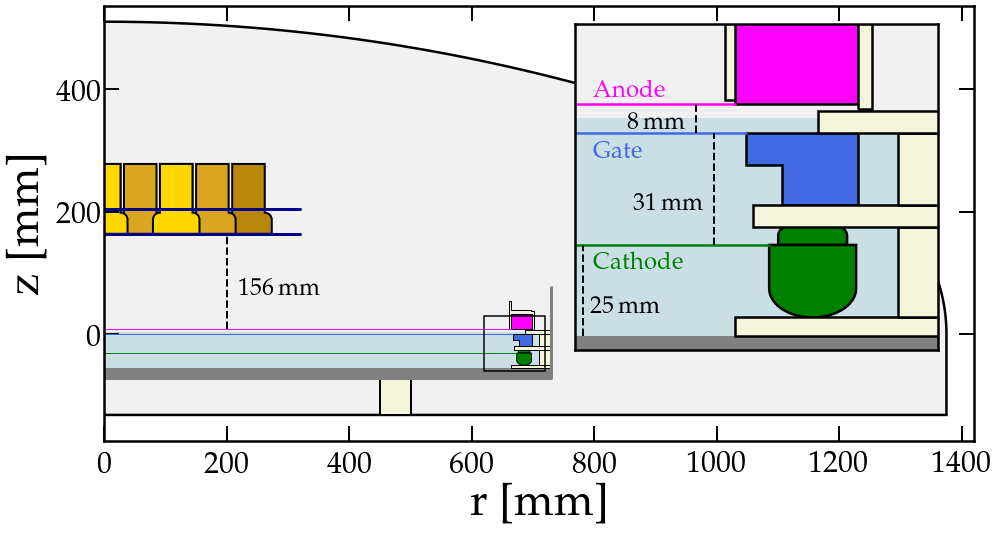}
    \caption{Schematic side view of the inner cryostat with the installed PMT array (gold) and TPC. The inset depicts the electrode distances; the colored horizontal lines indicate the electrode planes. The uncertainty on the distance between the anode plane and the PMT array is around 10\,mm. }
    \label{fig:side_schematic_jaron}
\end{figure}

The shape of the electrode frames and the $\sim$\SI{400}{\kg} of available xenon defined the lateral dimensions of the TPC, 
which was installed in a cylindrical open-topped vessel with a diameter of \SI{1460}{mm} and a height of \qty{150}{cm} to increase the LXe level around the TPC.
The vessel was made of \qty{2}{mm} thick stainless steel, and supported by six PTFE standoffs attached to the bottom of the vessel. A second "false" floor made of a \qty{3}{mm}-thick, mirror-polished stainless steel sheet with a diameter of $\sim$\SI{146}{cm} was installed inside the vessel to create a flat surface below the cathode. 
The vertical position of the cathode plane, 25\,mm above the vessel floor, was chosen  based on finite-element field simulations: This keeps the electric field at the cathode frame well below 30\,kV/cm, even for cathode bias voltages around 12\,kV. The orientation of the cathode frame was chosen to avoid that the thin mesh electrode gets too close to the grounded vessel floor. The TPC length was defined by the \SI{31}{\mm} distance between the cathode and gate, see Fig.~\ref{fig:side_schematic_jaron}. When operated at \SI{2}{\bar}, the LXe filling height above the floor was \SI{60}{\mm}, defining the target liquid-gas interface \SI{4}{mm} above the nominal plane of the gate wires.
The distance between the anode and gate electrode planes was \SI{8}{\mm}, with the wires facing each other. No system to precisely adjust the LXe level was installed, which was thus defined by the amount of Xe inside the system and the gas pressure. With the anode and gate set to +\SI{1}{\kilo\volt} and \SI{-5.1}{\kilo \volt}, respectively, and a cathode potential of $-$5.5\,kV close to the gate potential, a drift field of \SI{140}{\voltspercm} was achieved, as verified by detailed COMSOL-based simulations. The corresponding extraction fields were \SI{3.7}{kV/cm} in liquid and \SI{7.2}{kV/cm} in gaseous xenon, assuming the liquid-gas interface at \SI{4}{mm} above the gate. High-voltage was provided to each electrode by Kapton-insulated  cables.  

\begin{figure}[h!]
    \centering
    \includegraphics[trim = 0 150 0 0, clip, width=\linewidth]{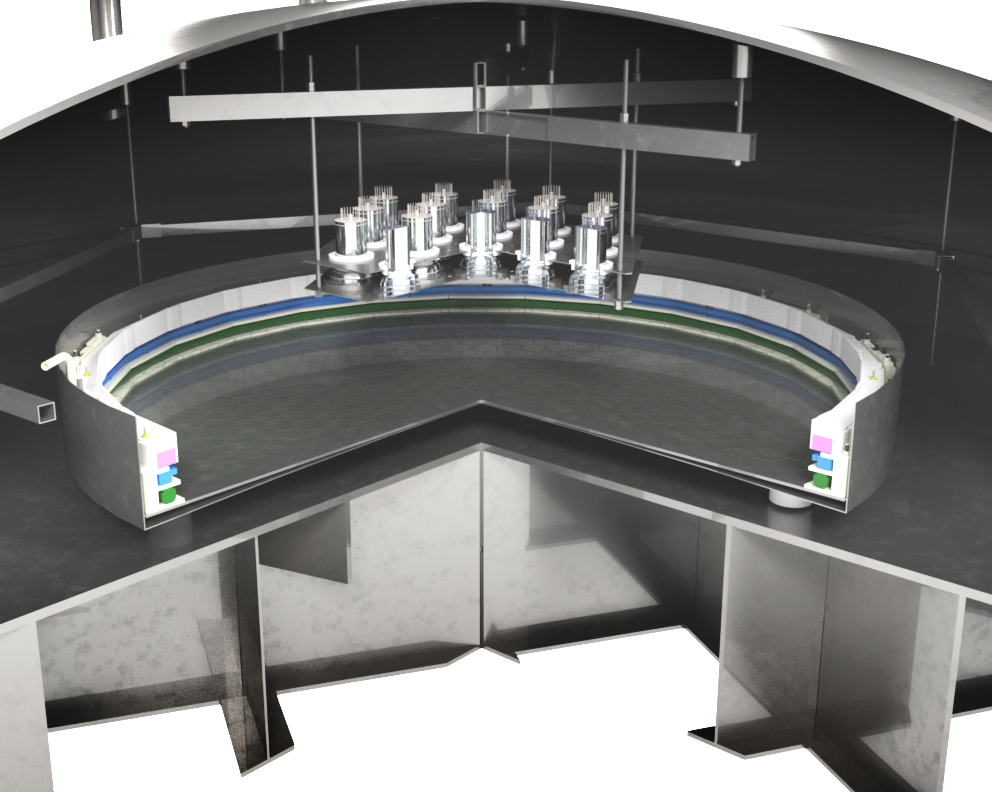}
    \caption{CAD rendering of the TPC setup inside PANCAKE. It was installed inside a cylindrical open-topped vessel of \SI{1460}{\mm} diameter which contained the entire LXe for the run. PTFE components were used to built a shallow TPC from the three electrodes, cathode (green), gate (blue), and anode (purple). Only the top of the anode was covered with PTFE reflectors. The PMT array was located centrally above the shallow TPC.  }
    \label{fig:CAD_rendering_electrodes_inside}
\end{figure}

Before TPC assembly in the ISO-6 cleanroom, the tensions of the anode and gate wires were measured independently (see Sect.~\ref{sec::wiretension}). 
PTFE spacers and support structures were used to build the TPC and to mitigate electric breakdown between the electrode frames due to high electric fields. Additionally, \SI{75}{\micro \m} thick Kapton foil was used in several positions to avoid direct line of sight between the electrode frames and the lateral wall of the open-topped vessel to mitigate electric breakdown. Wherever precise definition of distances was required, e.g., to ensure the \SI{8}{\mm} distance between gate and anode, Torlon or PEEK spacers were utilized. The anode frame was embedded in PTFE pieces, however, the design was generally not optimized for light reflectivity at the TPC edges since the small PMT array was installed above the center part of the TPC, see Fig.~\ref{fig:CAD_rendering_electrodes_inside}. 

Four capacitive levelmeters, as described in~\cite{Brown:2023vgf}, were installed around the LXe target at a fixed height with respect to the gate/anode system to measure the LXe level. Based on their reading, the entire PANCAKE platform was leveled.

A LXe extraction pipe as well as a funnel to collect condensed Xe from the condenser and the heat exchanger were installed in the open-topped vessel for Xe purification. A total of 15~PT100 temperature sensors were installed around the TPC. Three sensors were placed at the heights of the cathode, gate, and anode electrodes around the TPC at four azimuthal angles to closely monitor their temperatures and temperature differences. Additional instrumentation, in particular the array of photomultiplier tubes (PMTs) installed about \SI{150}{\mm} above the anode plane, are described in the next section.

%%%%%%%%%%%%%%%%%%%%%%%%%%%%%%%%%%%%%%%%%%%%%%%%%%%%%%%%%%
\section{Instrumentation}
\label{sec::tools}

Various diagnosis tools were installed to understand the performance of the shallow TPC, with a focus on identifying possible high-voltage issues as well as studying the performance of such a large TPC in an unshielded environment. VUV light from interactions in xenon but also expected from discharges can be either detected by VUV-sensitive PMTs (see Sect.~\ref{sec:inst_pmt_array_daq}) or digital cameras that are also sensitive in the VUV range (see Sect.~\ref{sec::camera}). While the TPC electrodes were biased, the bias voltages and currents provided by the high-voltage system (see Sect.~\ref{sec:hvsyst}) were monitored by the slow control system. Additional instrumentation included a plastic-scintillator-based muon telescope (see Sect.~\ref{sec::muon}) and a device to precisely measure the tension of the electrode wires (see Sect.~\ref{sec::wiretension}), albeit not while they were installed in the TPC. 

%%%%%%%%%%%%%%%%%%%%%%%%%%%%%%%%%%%%%%%%%%%%%%%%%%%%%
\subsection{PMT array and data acquisition}
\label{sec:inst_pmt_array_daq} 

For single-photon detection and TPC characterization, one array of 19~Hamamatsu R11410-21 photomultiplier tubes (PMTs) with a diameter of 3 inches~\cite{XENON:2015ara} was suspended centrally from the inner cryostat lid in the Xe gas phase, as shown in Fig.~\ref{fig:CAD_rendering_electrodes_inside}. 
The PMTs were arranged in a hexagonal structure with a distance of 102\,mm between the PMT centers and were fixed to two \SI{3}{\mm}-thick stainless steel plates using individual PTFE discs for electrical insulation. High-voltage was provided to each PMT individually from a CAEN A1535N power supply via Accu-Glass Kapton insulated wires. The PMT signals were read by RG196 A/U coaxial cables with MMCX connectors. Three blue LEDs were placed in different positions inside the inner cryostat for PMT gain calibrations using an Agilent 33220A pulse generator. Results are presented later in Sect.~\ref{sec::pmtcal}.

\begin{figure}
    \centering
    \includegraphics[width=\linewidth]{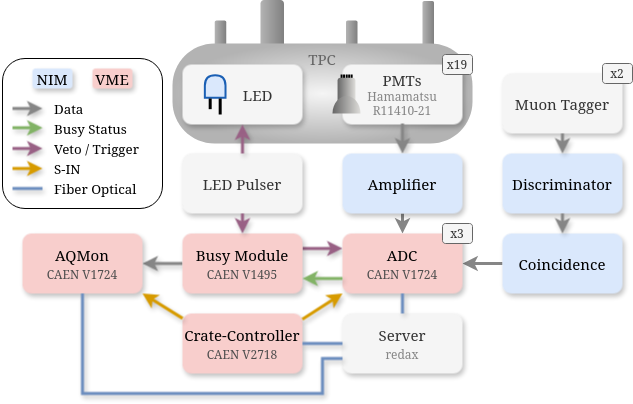}
    \caption{Block diagram of the hardware in the PANCAKE DAQ system: The~19 amplified PMT signals were digitized by three 8-channel waveform digitizers. Coincident muon telescope signals above the discriminator threshold produced a logic signal that was digitized by one the ADC channels. 
    For PMT calibration, a pulser drives an LED and triggers all ADCs, where the trigger signal is distributed by the Busy model which, in standard TPC operation, collects Busy signals from individual ADCs for further distribution. 
    The ADCs are read out by a server via optical fibers.}
    \label{fig:daq_diagram}
\end{figure}

The data acquisition (DAQ) system is largely based on the system that was successfully developed for the XENONnT experiment~\cite{XENON:2022vye}. The absence of a global trigger allowed data-taking in the unshielded environment by accepting exposure losses due to deadtime and by increasing the self-trigger threshold for each channel by a factor six compared to XENONnT, which still digitizes 80\% of the single photoelectron distribution.  
The DAQ system used here is illustrated in Fig.~\ref{fig:daq_diagram}: The 19~PMT signals were first amplified by a factor of 10 and then digitized by three 8-channel CAEN V1724 sampling digitizers, providing a sampling rate of 100\,MHz at 14-bit resolution. The ADCs were time synchronized using a common clock and a synchronous acquisition-start signal issued by the crate controller shown in Fig.~\ref{fig:daq_diagram}. During standard operation, the DAQ was run in self-trigger mode, recording signals exceeding a configurable self-trigger threshold independently for each channel. The information from all channels was combined in software based on their timestamps. A spare ADC channel recorded the logical muon telescope (see Sect.~\ref{sec::muon}) signal to allow for precise timing of tagged muons. Data processing is described in Sect.~\ref{sec::results}.

The busy signals from the ADCs are collected by a CAEN V1495 programmable digital multi-purpose module (Busy Module), which issues a veto to all ADCs. Its custom firmware also features a periodic "fractional lifetime" mode that enables data-taking only for a short time intervals. This allows data-taking in very high-rate conditions avoiding incomplete data collection due to busy situations, whilst maintaining a high single-photon detection efficiency. This is an alternative to raising thresholds on the channels. During LED calibration of the PMTs, the ADCs are externally triggered by a pulser via the Busy Module. The same pulser also directly drives the LEDs inside the TPC.

%%%%%%%%%%%%%%%%%%%%%%%%%%%%%%%%%%%%%%%%%%%%%%%%%%%%%
\subsection{Cryogenic cameras} 
\label{sec::camera}

\begin{figure}[t]
    \centering
    \begin{subfigure}[b]{0.5\textwidth}
        \centering
        \includegraphics[width=\linewidth]{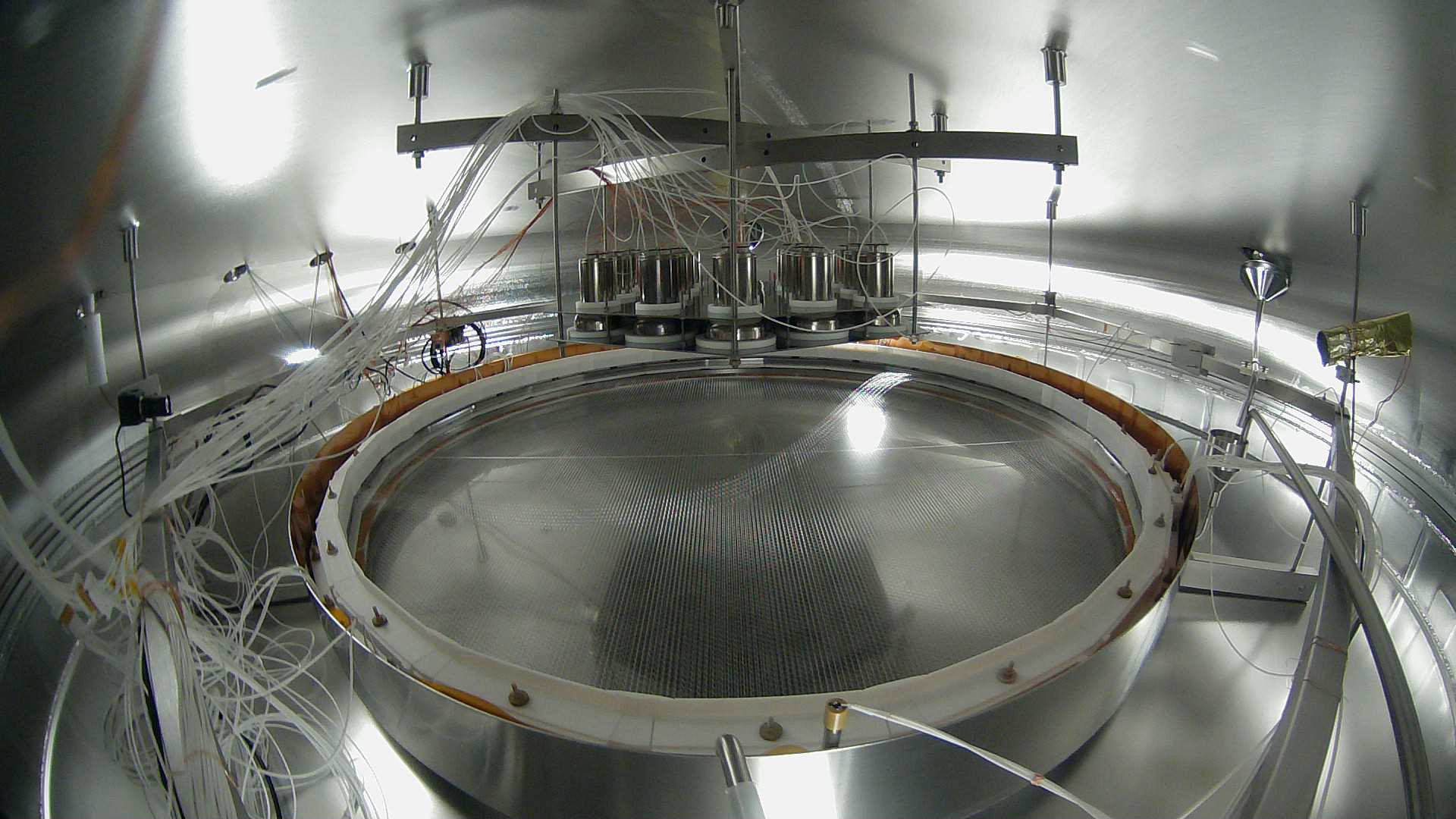}
    \end{subfigure}
   % \vspace{0.5cm}
    \begin{subfigure}[b]{0.5\textwidth}
        \centering
        \includegraphics[width=\linewidth]{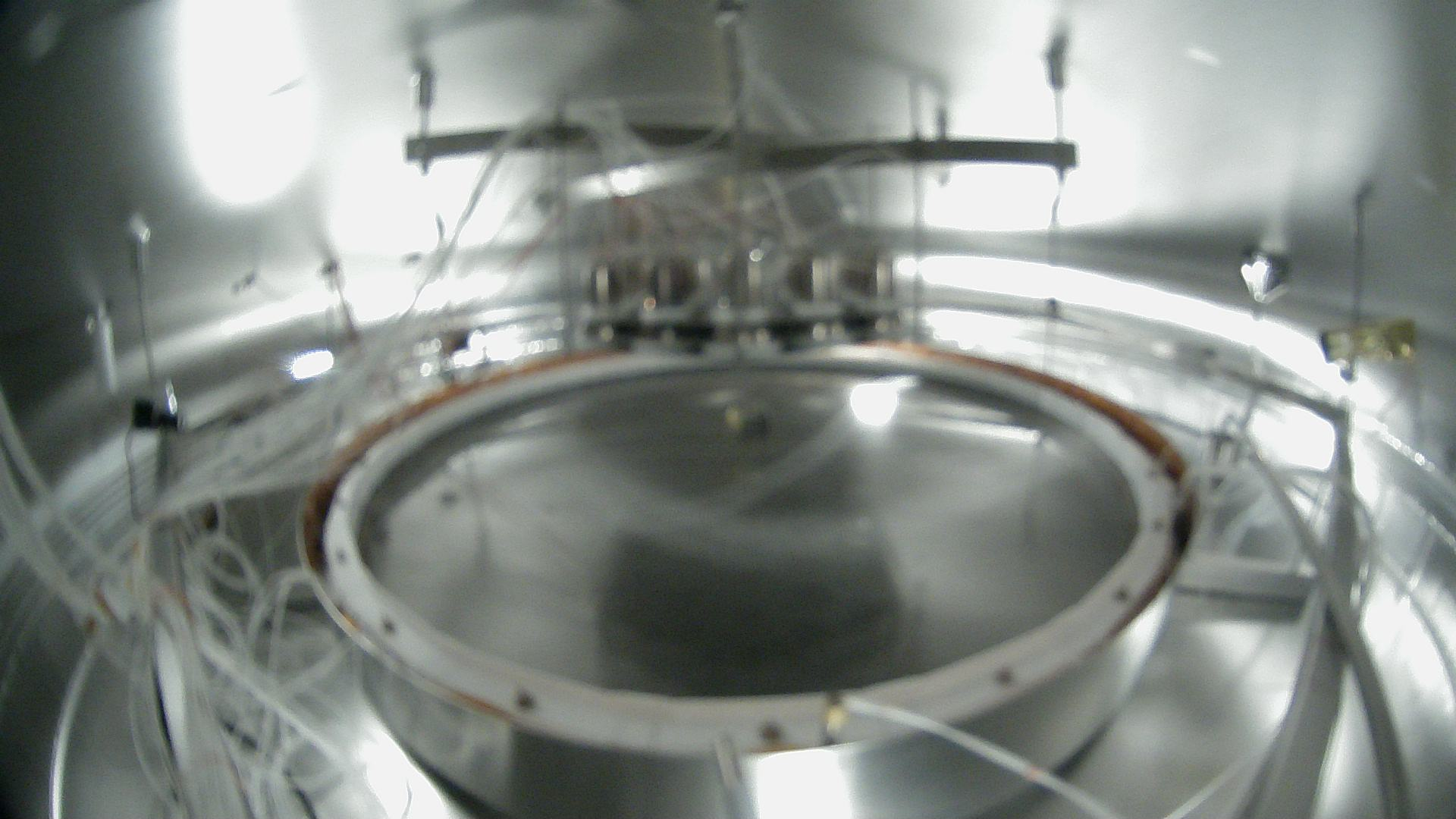}
    \end{subfigure}
    \caption{Inner cryostat illuminated by the internal LEDs. The images were captured by the overview camera when the cryostat was at room temperature (top) and when it was cold and filled with liquid xenon (bottom). The images show that the camera loses focus when being cold. This is likely due to thermal shrinkage and is reverted once the camera is at room temperature again. }
    \label{fig:fishy_frame}
\end{figure}

\begin{figure}[t]
    \centering
\includegraphics[width=\linewidth]{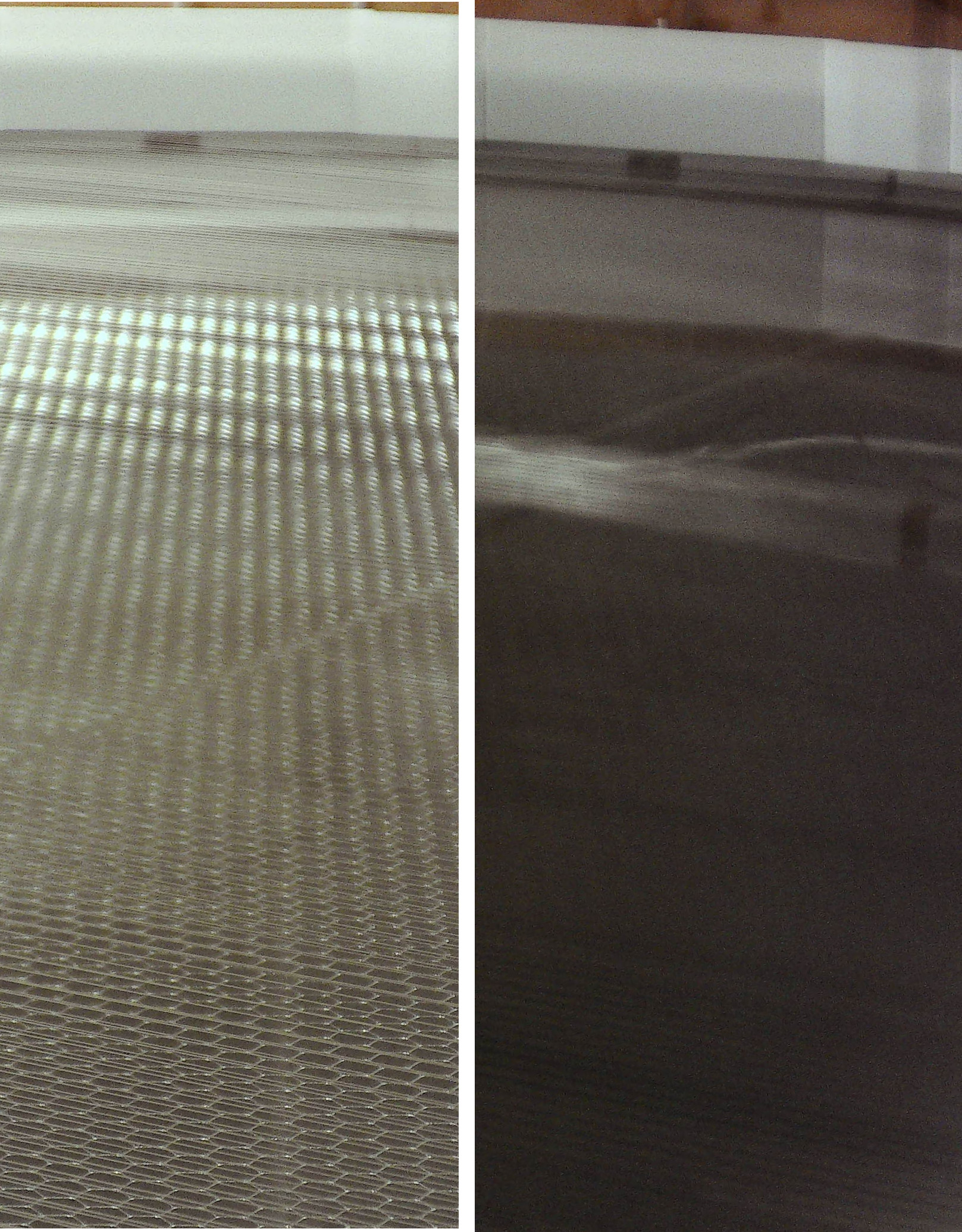}
    \caption{Two images of the grid camera taken at cryogenic temperatures. The increasing LXe level from the left to the right image obstructs the view onto the hexagonal cathode mesh. 
    Individual electrode wires, as well as the weld seam on the cathode running from the bottom left corner to the top right corner across both images, can be resolved.    }
    \label{fig:grid_cam}
\end{figure}

Standard USB webcams can operate at cryogenic temperatures when they are being heated~\cite{Brown:2023vgf}, and three such cameras were operated inside the inner vessel. The ELP 0.01 Lux USB-2.0 camera (overview camera) with fisheye optics provides an overview image of the inner cryostat and the TPC test setup inside. The image quality worsened during Xe operation, most likely due to thermal shrinkage in the optical lens system of the camera, but remained sufficient to see smaller components, see Fig.~\ref{fig:fishy_frame}. Additionally, a Svpro 8.0MP USB camera module, equipped with a 5-\SI{50}{\mm} CS mount optical lens and 10$\times$ zoom, was used (grid camera). This camera allowed individual gate and anode wires to be resolved and even enabled observation of the electrostatic sag with biased electrodes at cryogenic temperatures, see Fig.~\ref{fig:grid_cam}. An XLayer USB webcam was installed as well but was not used further for the work presented here. In the cryostat the cameras were connected via active USB cables which were connected to Sub~D-15 or multi-pin feedthroughs.   

Initially, tests to gradually increase the high-voltage were performed while recording videos with one camera, which were then manually analyzed to identify light emission in the footage. 
Later operations involved automatic, real-time monitoring of all three cameras using a custom-developed program that processed the camera feed in real time with OpenCV2 on a dedicated Linux system. This package also allowed for increasing the frame exposure time of the camera sensor, which increased the sensitivity of these cameras to low light signal. The program processed the feed frame by frame and only saved a frame if a light signal was detected during the TPC operation of the detector, which increased storage efficiency. 

%%%%%%%%%%%%%%%%%%%%%%%%%%%%%%%%%%%%%%%%%%%%%%%%%%%%%%%
\subsection{High-voltage system} 
\label{sec:hvsyst}

A CAEN N1470 module providing a maximum of $\pm$\SI{8}{\kilo\volt} and a current resolution of \SI{5}{\nano\ampere} was used to deliver high voltage to the anode and the gate electrodes. The current was monitored every second by the slow control system. 
The electrodes were connected to SHV coaxial feedthroughs via multi-stranded, Kapton-insulated HV cables from Accu-Glass. These cables had a \qty{10}{kV} DC rating and custom-made copper connectors. 

Two different high-voltage modules were used to bias the cathode during the run. A Heinzinger 60000-3ump provided a maximum voltage of \SI{60}{\kilo \volt} at \qty{20}{\micro \ampere} current resolution. Alternatively, a CAEN N570 module with a maximum of \SI{15}{\kilo \volt} was used. Only the current readings from the Heinzinger module could be read by the slow control system. 
Inside the cryostat, the cathode was connected to a \qty{30}{kV}~DC-rated, multi-stranded, Kapton-insulated, high-voltage cable from Accu-Glass. The cable was then connected to a 40\;kV-rated MDC HV40-1S-C40 ceramic feedthrough via a PTFE-insulated copper conductor.

%%%%%%%%%%%%%%%%%%%%%%%%%%%%%%%%%%%%%%%%%%%%%%%%%%%%%%%
\subsection{Wire-tension measurement system}  
\label{sec::wiretension}

The tension of the TPC electrode wires could be measured in the clean room using a system that determines the wire tensions by measuring the harmonic vibration modes of the wires. The system was developed based on the ideas described in~\cite{Prall:2010}. The device enables the tension of individual wires to be monitored during electrode construction and before or after delicate operations. Here, it was used to check the mechanical integrity of the wire electrodes after TPC operation in liquid xenon.

The oscillation of individual wires was induced by a short jet of pressurized air, nitrogen, or argon gas. A red class-2 laser was positioned above the wire and pointed at it at a slight angle. A photodiode mounted next to the laser observes the reflected light, whose intensity is modulated due to the oscillating wire. The photodiode signal is pre-amplified on a custom PCB board and digitized by an off-the-shelf Sound Blaster Play!\,3 USB sound card from Creative Technology Ltd.  The digitized signal undergoes a fast Fourier transform and background subtraction using a representative background spectrum. An algorithm for peak detection then analyzes the signal and identifies the normal mode of the wire based on its amplitude and distance from higher modes.  
The tension of an individual wire is computed from the frequency of the normal mode, the wire's length, diameter, and the density and Young's modulus of the wire material. Due to the finite measurement duration of the oscillating wire and the resolution of the digitizer, the frequency resolution is around \qty{0.3}{\Hz}, which translates into an uncertainty in the wire tension of less than \qty{0.2}{\N} for the wires used in this TPC.

The gas nozzle, laser, and photodiode were mounted on a stainless steel carriage that moved along a rectangular stainless steel profile using a stepper motor at one end of the profile and a timing belt. 
An energy chain guided the signal and power cables, as well as the gas hose, from the motor's side to the carriage.
Flexible supports at the ends of the profile allowed for flexible positioning above the investigated wire electrodes. The system was semi-automated, and a full electrode could be examined in about three hours. 

\begin{figure}
    \centering
\includegraphics[width=\linewidth]{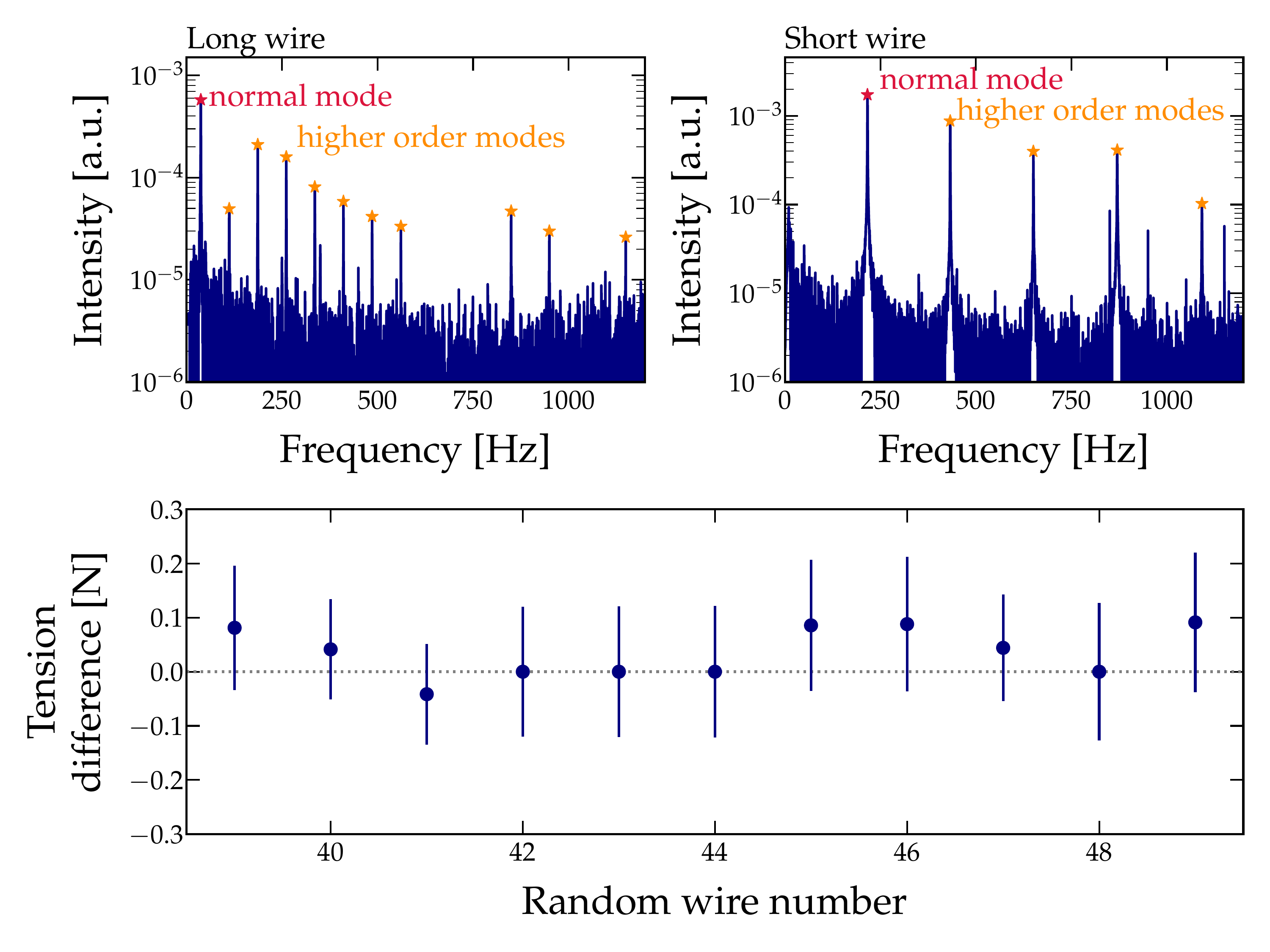}
    \caption{\textbf{Top:} Two frequency spectra for a long and a short wire, respectively, the normal and higher order modes are indicated. \textbf{Bottom:} Tension differences for a representative set of wires comparing the tension before and after the operation in liquid xenon. No significant tension change was observed for any of the wires.}
    \label{fig:witemeas_example}
\end{figure}

Figure~\ref{fig:witemeas_example} shows frequency spectra for a long and a short electrode wire. In addition to the normal mode, both spectra exhibit higher-order modes at multiples of the normal mode. Due to their similar tension of a few Newtons and significantly different lengths, the normal mode of the long wire is at a lower frequency than that of the short wire. The figure also illustrates the tension differences measured for a representative set of wires before and after operation in PANCAKE. These differences agree with zero within the measurement uncertainty, suggesting no loss in tension.

%%%%%%%%%%%%%%%%%%%%%%%%%%%%%%%%%%%%%%%%%%%%%%%%%%%%%%%%%
\subsection{Muon telescope}
\label{sec::muon}

A muon telescope consisting of two plastic scintillator panels could be installed below the PANCAKE facility. The panels of $10 \times 25$\,cm$^2$ and $10 \times 10$\,cm$^2$, respectively, were fixed at a relative distance of 70\,cm and each connected to a PMT. In principle, the muon telescope can be rotated to change the zenith angle; however, for the data presented here, it was only installed vertically at several different positions across the cryostat diameter.  
It was used to tag events in which a muon passed through the detector to evaluate the detector response for events with known interaction position.

The discriminated PMT signals were fed into a coincidence unit to create a muon tag; this logical signal was digitized together with the PMTs in the xenon volume ensuring proper time synchronization, see Sect.~\ref{sec:inst_pmt_array_daq}.

%%%%%%%%%%%%%%%%%%%%%%%%%%%%%%%%%%%%%%%%%%%%%%%%%%%%%%%%%%
\section{Operation} 
\label{sec::operation}

\begin{figure}
    \centering
    \includegraphics[width=\linewidth, trim = 0 200 100 220, clip]{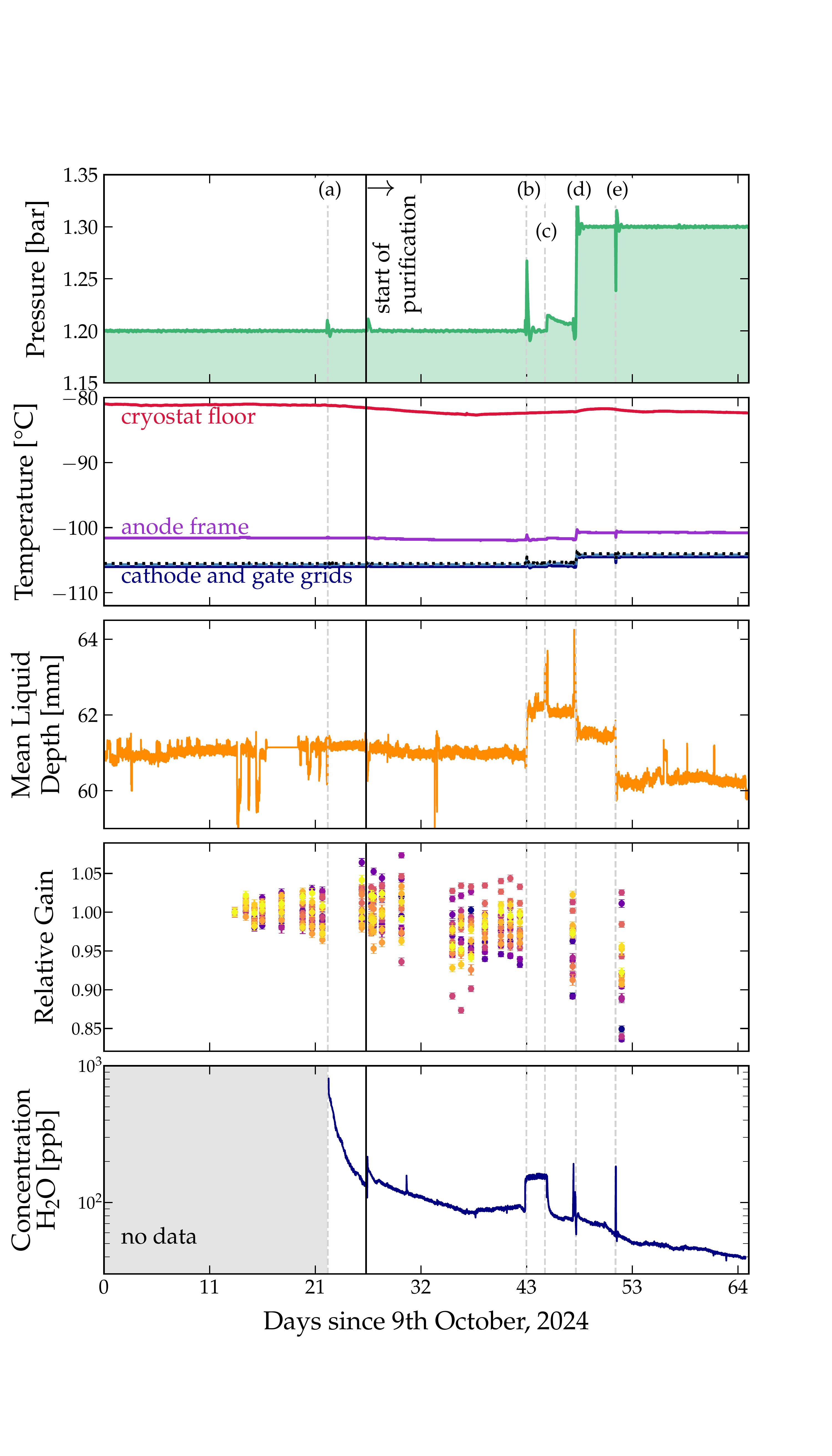}
    \caption{Operation parameters during the TPC phase. Major operations are indicated by the vertical lines: a) Test of purification system and HALO+, b) Filling additional xenon, c) Power outage, d) Change operation pressure from 1.2 to \SI{1.3}{bar}, e) Recuperation of xenon. The start of the xenon purification is indicated by the solid black line. No Xe purification between b) and c). \textbf{Panel~1:} Xenon gas pressure in the inner cryostat. Two different pressures were set to evaluate the impact. \textbf{Panel~2:} Temperatures measured at the TPC electrodes and the cryostat floor. The cryostat floor was warmer since it was not in direct contact with liquid xenon. \textbf{Panel~3:} Average value of all four levelmeters after detector leveling. %The larger spikes indicated by the dashed lines were induced by detector operations such as filling additional xenon, increased purification rates, gas pressure increase or recuperation of xenon.
    \textbf{Panel~4:} The relative PMT gains normalized to day~1 indicate that the gains were stable within $\pm$10\%. The colors represent the different PMTs. \textbf{Panel~5:} H$_2$O concentration measured with the HALO+ trace gas analyzer.} 
    \label{fig:run_overview_plot}
\end{figure}

The PANCAKE platform was operated for about four months at the end of 2024, with approximately 60~days of TPC operation. Figure~\ref{fig:run_overview_plot} shows several slow control parameters, such as as xenon gas pressure, temperatures, and the LXe level during the TPC phase. The platform was operated at xenon pressures of \SI{1.2}{\bar} and \SI{1.3}{\bar} to evaluate the impact of pressure on high-voltage stability. The xenon purification system was started after approximately 26\,days. The improving xenon purity was monitored using the HALO+ sensor, which is sensitive to H$_2$O, as well as TPC data (see Sect.~\ref{sec:purity}).

The inner cryostat vessel was initially cooled down by running liquid nitrogen through the six cooling pads attached to the inner cryostat until floor temperatures of $\sim$\SI{-100}{\celsius} were reached. The temperatures around the TPC electrodes were monitored by 15~PT100 temperature sensors. During operation,  maximal temperature differences of less than \SI{10}{K} across the entire TPC were observed. For each individual electrode, a maximal temperature difference of about \SI{5}{K} was found, keeping temperature gradients in a safe range given the different heat capacities of electrode frames and wires, which can lead to different thermal contractions.

Xenon was filled into the cryostat through the purification system with a typical gas flow of around \SI{2.7}{slpm}. The xenon condensed on the cold head of the thermosyphon cooling system and was guided into the open-topped vessel. The rise in liquid level was monitored by both the levelmeters and the grid camera, see Figs.~\ref{fig:grid_cam} and~\ref{fig:run_overview_plot}. 
The TPC was leveled by adjusting the three M48 support rods that support the cryostat before the final liquid level was reached; this was informed initially by grid camera images and later by the LXe levelmeter readings. 
Larger jumps and spikes in the levelmeter readings compromised the leveling procedure; however, the results presented later are not severely affected, since level difference was small across the area directly monitored by the PMTs. Improvements on the levelmeter response are essential for the next detector run. The mean filling height based on all four level sensors after leveling is shown in Fig.~\ref{fig:run_overview_plot}.

\begin{figure}[t]
    \centering
    \begin{subfigure}[b]{0.5\textwidth}
        \centering
        \includegraphics[width=\linewidth]{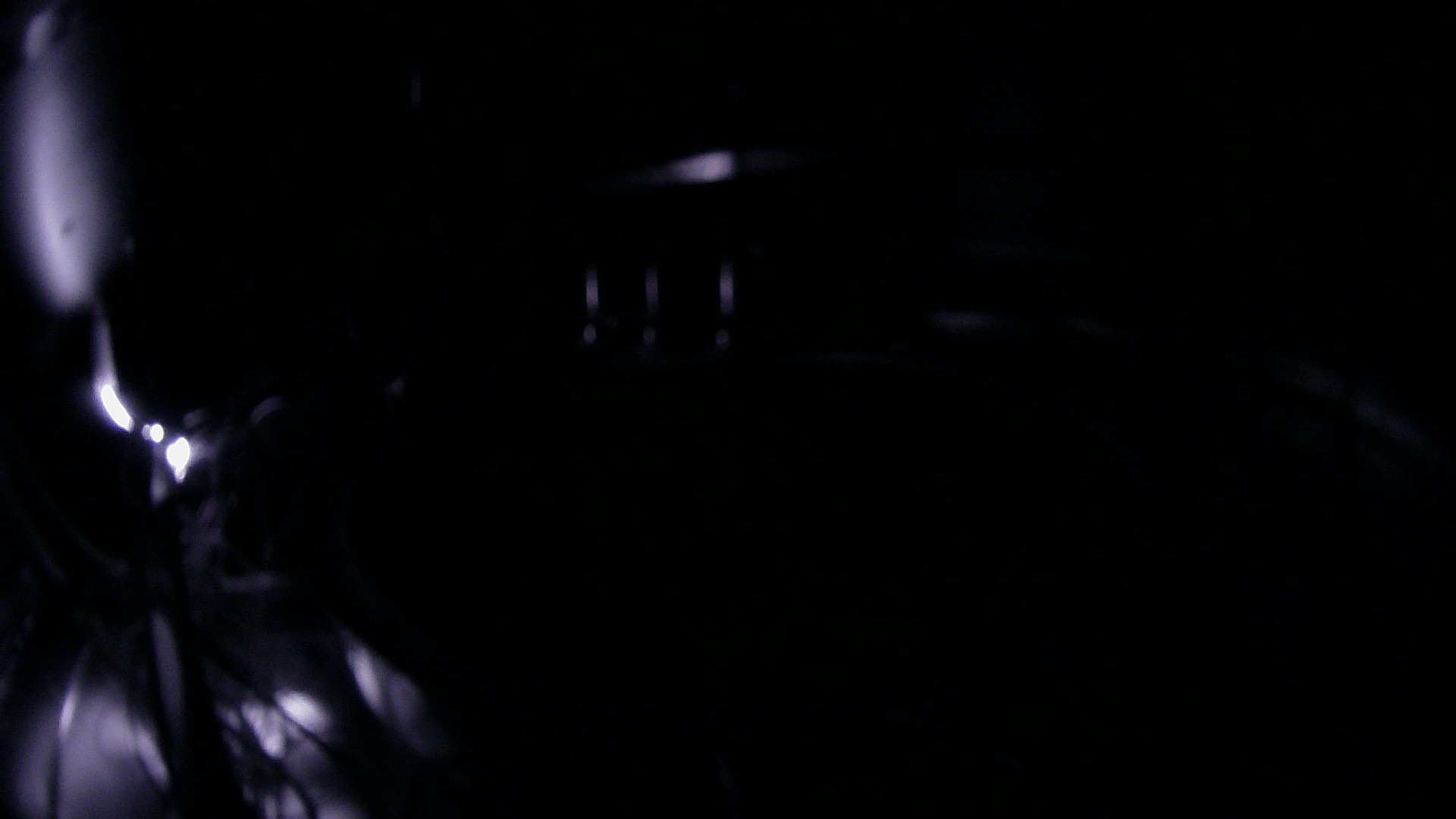}
    \end{subfigure}
   % \vspace{0.5cm}
    \begin{subfigure}[b]{0.5\textwidth}
        \centering
        \includegraphics[width=\linewidth]{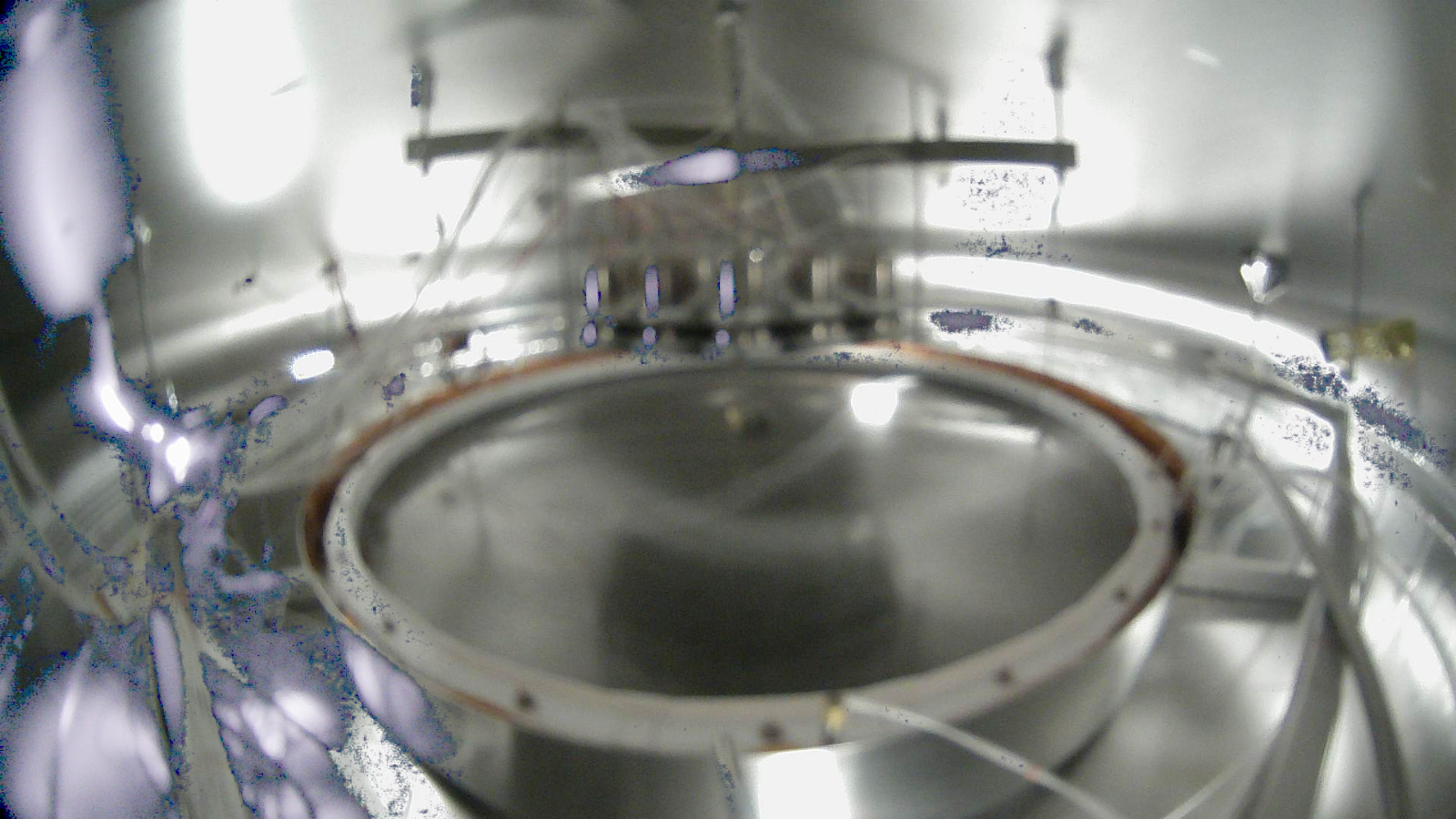}
        \label{fig:subfig2}
    \end{subfigure}
    \caption{\textbf{Top:} The camera image shows large light emission for very high cathode voltages. \textbf{Bottom:} Overlaying the top (dark) image  onto an illuminated reference image reveals that the light was emitted along the high voltage cable connected to the cathode and coiled up in the Xe gas phase (left). Reflections at the PMT array (top center) and the cryostat floor (bottom left) can also be seen. This indicates that the voltage range in this run was limited by the cable. }
    \label{fig:cathode_cable_glow_part}
\end{figure}

To evaluate the energy threshold of the TPC, \kr was injected into the purification gas flow. Data were acquired in different configurations, for example with varying ADC thresholds or TPC voltages. 
When biasing the cathode beyond \SI{-6}{\kilo \volt}, large light emissions could be observed by the cameras along the cable connecting the feedthrough to the cathode frame, as shown in Fig.~\ref{fig:cathode_cable_glow_part}. As similar events, although far less luminous, were observed around sections of anode and gate cables in the xenon gas volume, this suggests that the high-voltage range in this run  was limited by the installed cables.

%%%%%%%%%%%%%%%%%%%%%%%%%%%%%%%%%%%%%%%%%%%%%%%%%%%%%%%%%
\subsection{PMT calibration and performance} 
\label{sec::pmtcal}

The single photoelectron (SPE) response of all 19~PMTs was calibrated using one low-intensity, blue LED installed close to the PMT array.
The LED was driven by a signal generator to produce a small light signal which generated a PMT signal only \qty{10}{\percent} of the triggers, increasing the likelihood for single-photon detection.
Data were recorded in a \qty{2}{\micro\second} window around each trigger to allow for accurate estimation of the baseline signal before any LED-induced signal.
The spectrum of integrated LED signals was fit by a parametrized model of the SPE response to obtain the gain for each PMT from the mean of the SPE peak. 

The PMT voltages were adjusted individually to achieve a gain of \num{5e6} PE per photon. The PMT voltages were kept constant and the gains were regularly monitored. Figure~\ref{fig:run_overview_plot} (Panel~4) shows that the gains remained stable within \qty{10}{\percent} during TPC operation. 

The afterpulse rates of the PMTs, caused by residual gas in the tubes, were studied using the same pulsed LEDs with increased light output, according to the procedure described in~\cite{pmt_afterpulse_ref}. All PMTs exhibited afterpulses from helium and argon. We defined the afterpulse rate as the probability of a SPE producing a detectable afterpulse peak and found rates of \SIrange{0.1}{0.4}{\percent} for He$^+$ and Ar$^+$, while the rate for Ar$^{++}$ was below \qty{0.1}{\percent}. 
The signal size of these afterpulses varied from a few tens to a few hundred PE; however, the average remained below three PE, which did not impact the subsequent analysis.

%%%%%%%%%%%%%%%%%%%%%%%%%%%%%%%%%%%%%%%%%%%%%%%%%%%%%%%%%%
%%%%%%%%%%%%%%%%%%%%%%%%%%%%%%%%%%%%%%%%%%%%%%%%%%%%%%%%%%
%%%%%%%%%%%%%%%%%%%%%%%%%%%%%%%%%%%%%%%%%%%%%%%%%%%%%%%%%%
\section{Results} 
\label{sec::results}

Results from detector operation are presented here. The detector was operated in light-only mode, i.e., without biasing the electrodes, and in TPC mode. Background data from environmental radioactivity, mono-energetic conversion electrons from a \kr source and cosmic muons are used to characterize the performance. 

Here we focus on data acquired with the PMT array, since the cameras integrate their signals over microseconds and have a significantly higher detection threshold. Whenever the cameras detected a signal, the PMT waveforms were fully saturated. 

The digitized PMT data were processed using the analysis framework \emph{strax}~\cite{strax} together with an experiment-specific extension, which is largely based on the XENONnT processing chain \emph{straxen}~\cite{straxen} but uses parameters adapted to the PANCAKE detector.
After baseline subtraction, a PMT \emph{hit} is registered when the waveform exceeds a fixed threshold of 50~ADC counts. Hits from different channels are grouped into intermediate objects called \emph{peaklets} if their time separation is smaller than \SI{80}{\nano\second}, while isolated hits with larger separations are stored separately.
Peaklets are further split using a natural-breaks algorithm~\cite{jenks_natural_breaks} that evaluates whether a waveform is better described as one or two segments.
They are then classified as S1-like, S2-like, or unclassified using the peaklet size and rise time. This time is defined as the interval between 10\% and 50\% of the cumulative signal, which is generally much shorter for S1s than for S2s.
Peaklets classified as S2-like are subsequently merged with nearby S2-like peaklets and isolated hits if their temporal separation is sufficiently small, in order to undo potential over-splitting of wide S2 signals. 
Finally, the S1-like peaklets and merged S2s are stored as \emph{peaks}, a lightweight data format containing the reconstructed pulse properties. Most important for the subsequent event reconstruction and analysis are the integrated signal (area), converted into photoelectrons (PE), and the width, defined as the time interval between 25\% and 75\% of the area.

\begin{figure}
    \centering    \includegraphics[width=\linewidth]{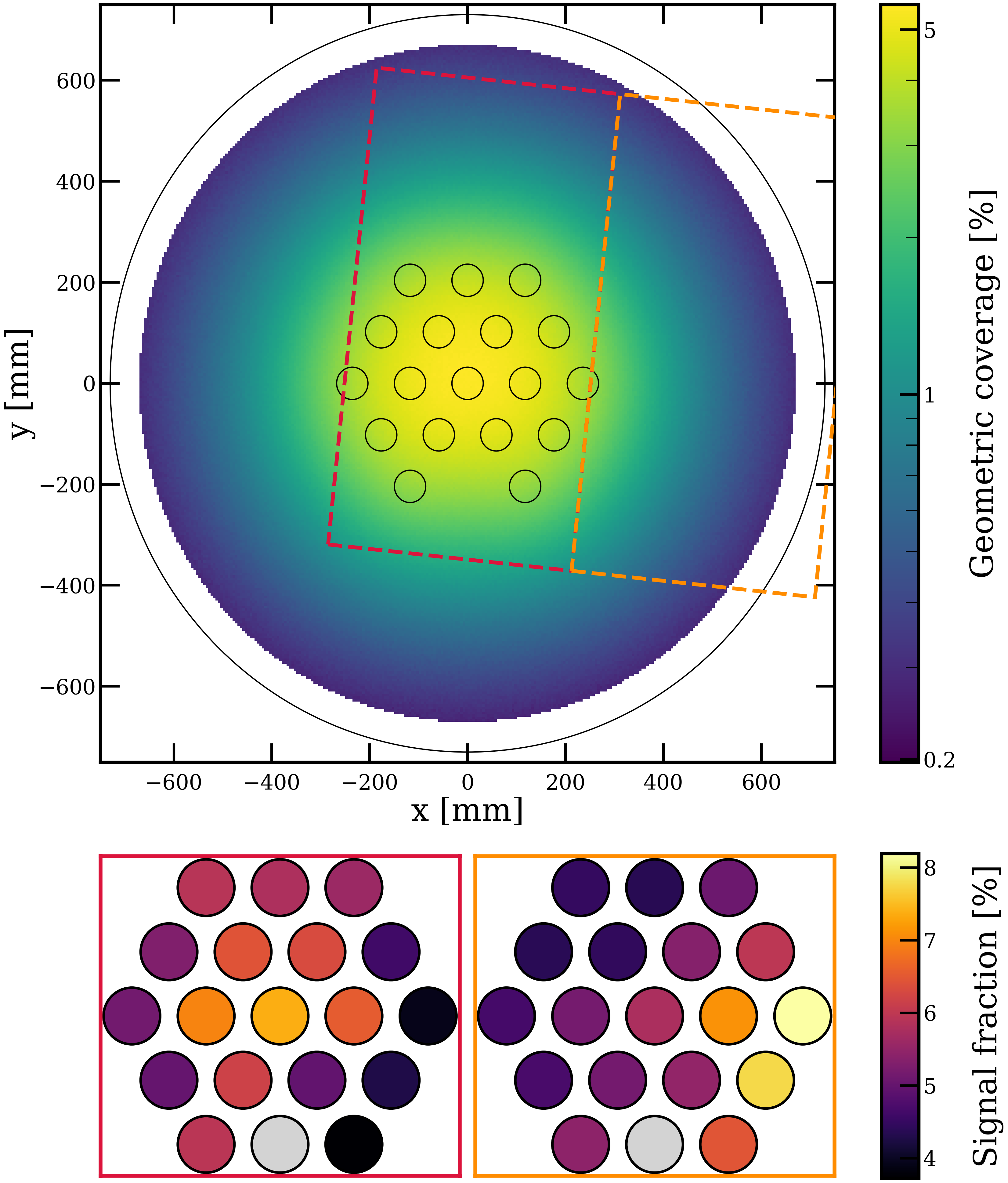}
    \caption{\textbf{Top: } Simulated geometric coverage of positions across the active TPC, averaged over the full LXe depth. The positions covered by the muon telescope are indicated by the colored boxes. The large circle indicates the open-topped vessel. \textbf{Bottom: } Average PMT hit patterns generated by muon-induced interactions in the LXe at these positions.  }
    \label{fig:muon_tagger_pos}
\end{figure}

%%%%%%%%%%%%%%%%%%%%%%%%%%%%%%%%%%%%%%%%%%%%%%%%%%%%%%%%%%
\subsection{Light-only analyses}
\label{sec:lightonly}

Here we present results from operating the detector without biasing the TPC electrodes. In this case, only S1~light signals are produced.
The absence of field quenching results in larger S1 signals than those produced in the presence of electric fields (TPC mode). 

%%%%%%%%%%%%%%%%%%%%%%%%%%%%%%%%%%%%%%%%%%%%%%%%%%%%%%%%%%
\subsubsection{Muon-induced events} 

As minimum ionizing particles (MIPs), cosmic muons deposit a fixed amount of energy per unit track length in liquid xenon. 
Muon-induced signals can therefore be used to assess the uniformity of the detector response. 
The muon telescope (Sect.~\ref{sec::muon}) was placed in various positions below the cryostat to tag muons when the detector was operated in light-only mode.
Coincident signals in both panels of the muon telescope generated a square pulse that was digitized by the DAQ system. This allowed the identification of PMT signals from the LXe detector in coincidence with tagged muon events. 
The data were collected using an increased digitization threshold of \qty{1000}{ADC} units per PMT channel to reduce the number of small LXe signals that dominate the rate. The low light collection efficiency in the outer detector regions and accidental coincidences induced a sizable fraction of muon-tagged events without any signal in the LXe detector. 

Figure~\ref{fig:muon_tagger_pos} shows the detector regions covered by two muon telescope positions: one covers the PMT array, while the second monitors the outer part of the LXe detector, with the center of the active area being about \qty{500}{\mm} from the central PMT. 
The two hit patterns shown in the bottom figure were generated by averaging all muon-tagged events at the two positions and clearly show the expected shift in the distribution of the PMT signals across the array. The 83\% reduction of the average muon signal from the inner to the outer muon-telescope position agrees very well with the expected $\sim$80\% reduction of the averaged geometric coverage, which is also shown in Fig.~\ref{fig:muon_tagger_pos}. 
The geometric coverage is defined as the position-dependent fraction of photons produced in the LXe hitting a PMT photocathode. It was simulated ignoring light reflections and using a simplified geometry, but integrates over the entire LXe depth. The systematic uncertainty on the simulation is about 40\%, dominated by the positioning uncertainty of the muon telescope.

%%%%%%%%%%%%%%%%%%%%%%%%%%%%%%%%%%%%%%%%%%%%%%%%%%%%%%%%%%
\subsubsection{Energy threshold }

Gaseous \kr released from a rubidium source was introduced into the detector by mixing it with the xenon gas in the purification system.
Due to its characteristic decay, which creates mono-energetic energy depositions at \qty{32.1}{keV} and \qty{9.4}{keV} in a fast delayed coincidence with a half-life of \qty{157}{ns},
\kr is a standard calibration source for LXe TPCs~\cite{Kastens:2009pa}.  
Due to the specifics of how the source was connected to the gas system and the dimensions of the piping, the calibration procedure required evacuating the pipes toward the LXe detector prior to opening the calibration volume containing krypton mixed with gaseous xenon.
About one minute after the valve was opened, a rate increase of approximately \qty{3}{\percent} above the background S1 rate of around \qty{9.3}{\kilo\Hz} was observed. 

\begin{figure}
    \centering
    \includegraphics[width=\linewidth]{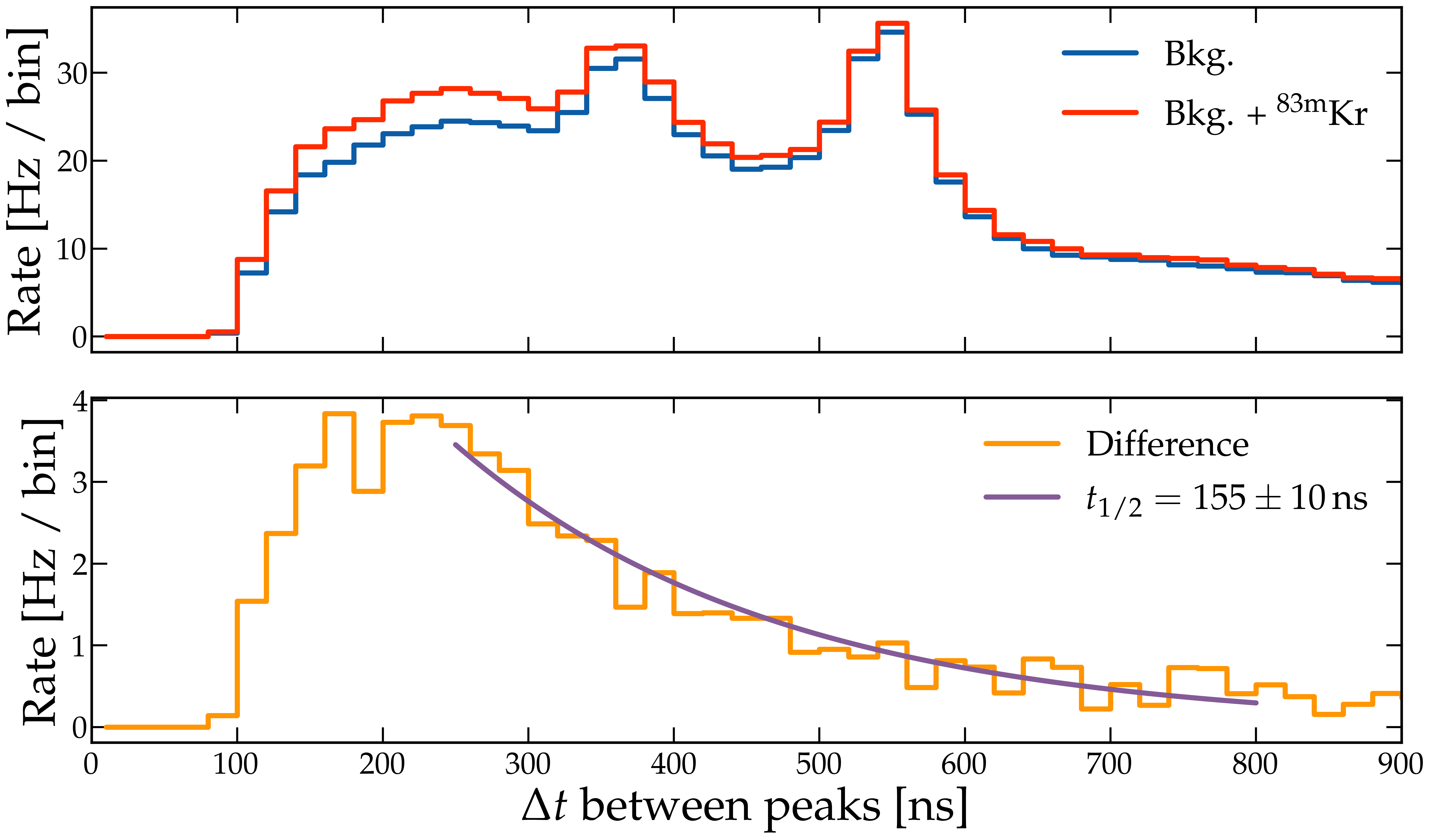}
    \caption{\textbf{Top:} Time differences between consecutive S1 peaks before (blue) and after \kr-injection (green). \textbf{Bottom:} Rate difference between both datasets. An exponential fit yields a half-life 
  of \SI{155\pm10}{\ns}, compatible with the literature value for the decay of the intermediate \kr state.}
    \label{fig:kr_time_diff}
\end{figure}

The dataset is separated into periods before and after this rate increase. Figure~\ref{fig:kr_time_diff} shows the time difference $\Delta t$ between consecutive S1~peaks. 
At time differences below \SI{900}{\ns}, one can see a clear excess in the data with \kr (green) with respect to the background-only data (blue). The difference, shown in the bottom panel of Fig.~\ref{fig:kr_time_diff}, shows the exponential shape of the excess. The fitted half-life of \SI{155\pm10}{\ns} is fully consistent with the literature value, confirming that these events were \kr decays. 
The typical time separation of independent consecutive events is around \SI{100}{\micro\second}. 

\begin{figure}
    \includegraphics[width=\linewidth]{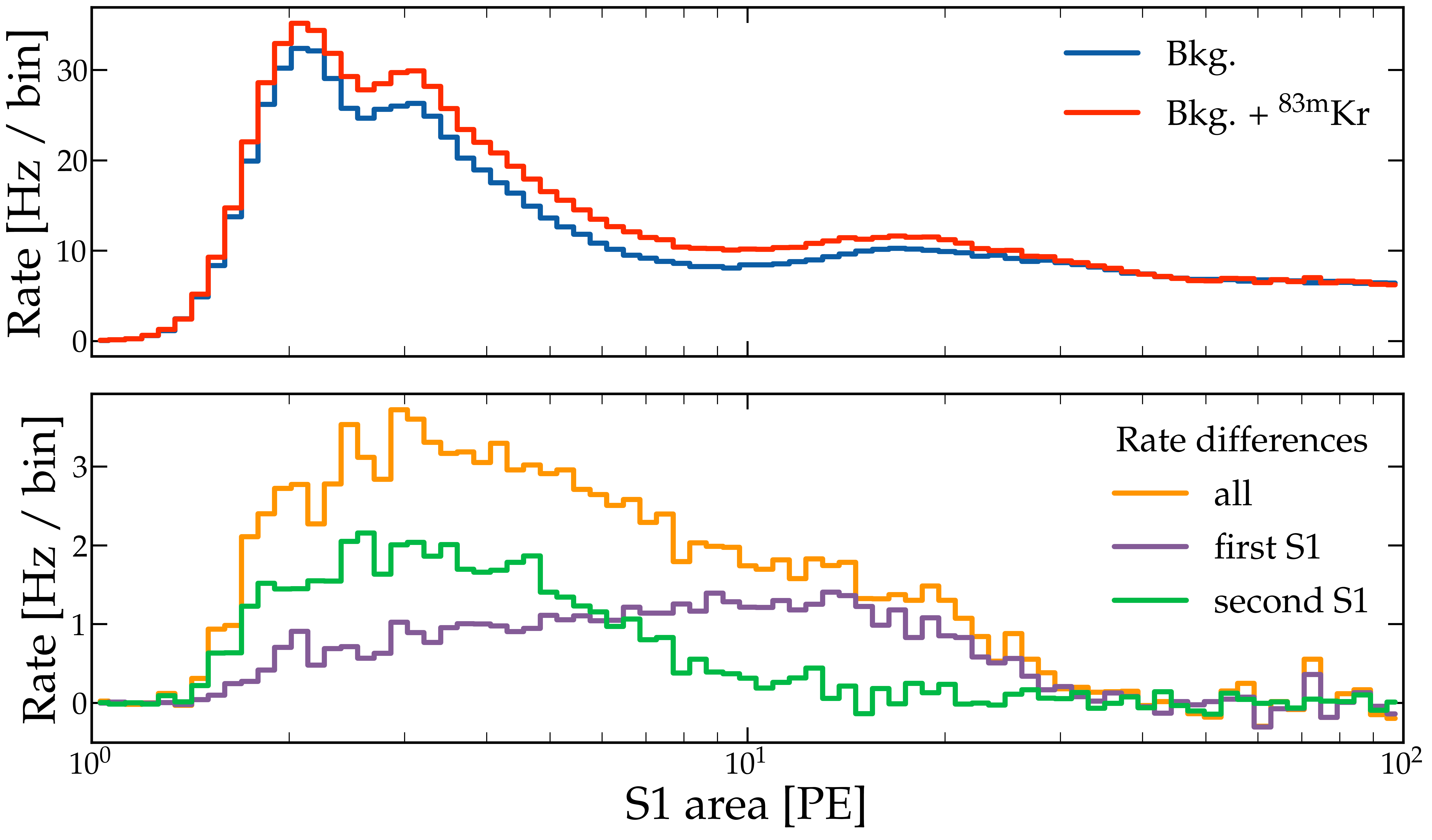}
    \caption{\textbf{Top:} Signal spectrum of all S1~events within time differences $\Delta t < \SI{800}{\ns}$. The increase from \kr events is clearly visible. \textbf{Bottom:} The difference spectrum from \kr (orange) is separated into the spectra of the time ordered first (purple) and second (green) S1 peak from the delayed coincidence decay of 32.1\,keV and 9.4\,keV signals, respectively. The 9.4\,keV signal extends down to the S1~threshold of 2\,PE.
    \label{fig:kr_area_diff}}
\end{figure}

Figure~\ref{fig:kr_area_diff} shows the corresponding signal area spectra after applying a cut requiring a second S1 peak within \SI{800}{\ns} before or after each peak. The excess of events compared to the background data without \kr is clearly visible for~S1 areas between the detection threshold at \SI{2}{PE} up to about \SI{30}{PE}. About \SI{88}{\percent} of peaks with $\Delta t < \SI{800}{\ns}$ come in pairs of exactly two, as expected for genuine \kr events. In fact, comparing the areas of the first and second S1 in these pairs, shown in purple and green in the lower panel of Fig~\ref{fig:kr_area_diff}, the results match the expected behavior of \kr where the first decay deposits \SI{32.1}{keV} and the second decay \SI{9.4}{keV}. The absence of narrow spectral features is understood as a consequence of the strong position dependence of the light collection in this geometry. 

These results indicate that the signal from the \SI{9.4}{keV} transition is well above the effective S1 detection threshold in the light-only campaign in the detector center.  
For calibration runs performed under TPC conditions, with drift and extraction fields applied, no clear \kr delayed coincidence signature could be identified. 
This is partly attributed to field-induced quenching of the scintillation light yield, which reduces the recombination probability of ionization electrons and thus the total S1 signal. 
A light yield reduction of about \SI{40}{\percent} is expected at drift fields around \SI{200}{V/cm}~\cite{quenching}, rendering a substantial fraction of the \SI{9.4}{keV} decays undetectable even in the center of the PANCAKE TPC with the photosensor array implemented for this work. 
Additionally, due to the presence of S2 signals, the background rate in the region of interest was found to be significantly higher under TPC conditions, further complicating the identification of \kr decays above the ambient background. 

%%%%%%%%%%%%%%%%%%%%%%%%%%%%%%%%%%%%%%%%%%%%%%%%%%%%%%%%%%
\subsection{TPC results}

In this section, we present results from operation periods in which all three electrodes were biased such that both prompt (S1) and delayed (S2) scintillation signals were generated. 

%%%%%%%%%%%%%%%%%%%%%%%%%%%%%%%%%%%%%%%%%%%%%%%%%%%%%%%%%%
\subsubsection{Event identification}
\label{sec:ident}

Prompt scintillation signals (S1s) were paired with the corresponding delayed proportional scintillation peaks (S2s) emerging from the same interaction. 
When operating in TPC mode without the \kr calibration source, the rate at which PMT signals exceeded the self-trigger threshold and occurred simultaneously across at least two PMTs was \qty{21}{\kilo\hertz}. 
This high rate can be attributed to the large active volume of the TPC and the absence of any shielding. The low light collection efficiency of the TPC at large radii moves many signals very close to the detection threshold.  
Therefore, several quality cuts were applied to achieve a cleaner event sample.
A time coincidence requirement, demanding signals in at least three PMTs within \SI{100}{ns}, effectively removes a substantial proportion of small, noise-like peaks while preserving genuine scintillation signals. 

The algorithm then iterates through a time-ordered list of the remaining S1 peaks and groups all additional S1s and all S2 peaks following within a time window of \SI{25}{\micro\second}, which is slightly longer than the expected maximal electron drift time, into an event candidate. In about \SI{80}{\percent} of cases, events contain exactly one S1 candidate. No S2 signals are found in \SI{42}{\percent} of the events, which correlates closely with the sizable mass fraction of xenon below the cathode from which no charges are extracted (\SI{25}{\mm} of the total LXe filling height of \SI{60}{\mm}).

\begin{figure}
    \centering    \includegraphics[width=\linewidth]{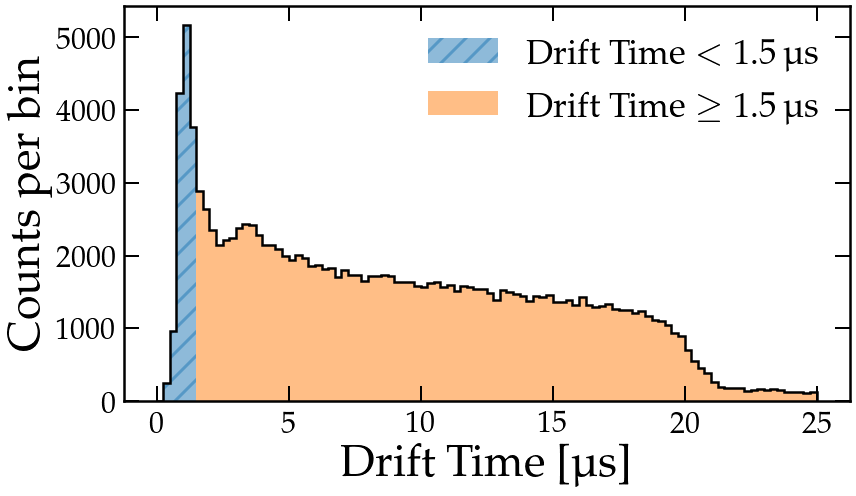}
    \caption{Electron drift time of events with exactly one S1 and one S2 peak at a drift field of \SI{140 \pm 10}{V/cm}. A cut is applied to remove events with drift times below \SI{1.5}{\micro\second} which are likely to be from gas events or random coincidences. The drop of events around \SI{20}{\micro\second} is due to the maximal drift time across the TPC. }
    \label{fig:drifttime_hist}
\end{figure}

Subsequently, events are required to contain exactly one S1 and one S2, and very short drift times $t_{\mathrm{drift}}=t_{\mathrm{S2}}-t_{\mathrm{S1}}<\SI{1.5}{\micro\second}$ are rejected to suppress signals produced in the gas region. This leads to the steep rise at small
$t_{\mathrm{drift}}$ in Fig.~\ref{fig:drifttime_hist}. The figure also shows that the algorithm identifies events up to a maximal drift time of about \SI{20}{\micro\second} after which only a small and constant number of accidental coincidences is found, clearly suggesting that the paired S1 and S2 peaks are indeed correlated.

\begin{figure}    
\centering\includegraphics[width=\linewidth]{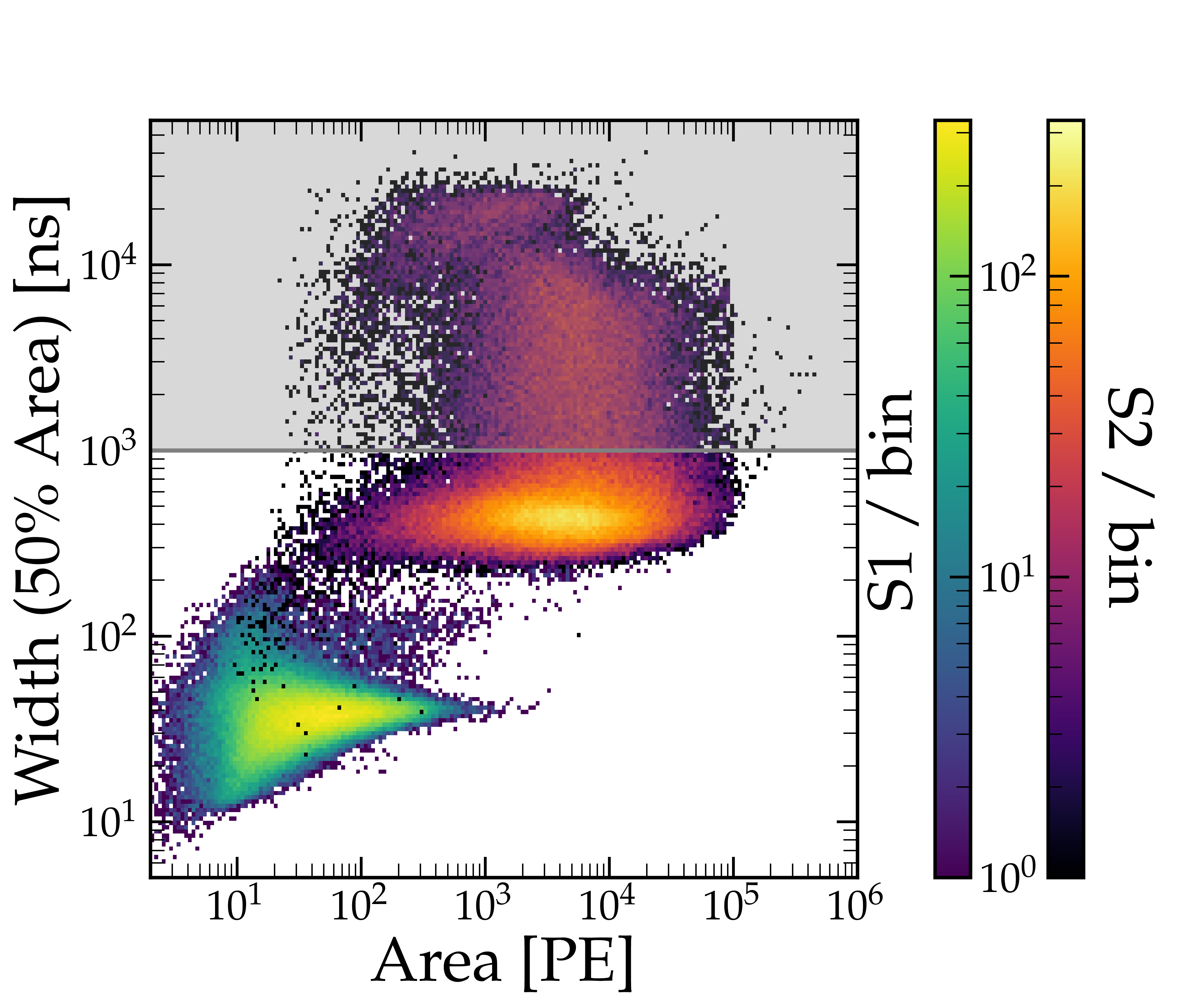}
    \caption{Distribution of signal width vs.\ area from S1 and S2 peaks from paired events. Events with S2 widths above \SI{1}{\micro\second} are rejected from the final event selection (gray area).}
    \label{fig:s1_s2_area_width}
\end{figure}

Figure~\ref{fig:s1_s2_area_width} shows that the S1 and S2 populations for the selected events are separated in the signal width vs.~area space. This is mainly caused by the different signal width, which is measured as the time interval around the peak maximum which contains 50\% of the signal area. S2 signals with widths above \SI{1}{\micro\second} were found to originate predominantly from incorrectly merged multi-scatter peaks. 
A fraction of these events is caused by tracks of atmospheric muons leading to signals with apparent widths of about \SI{10}{\micro\second} (i.e., \SI{50}{\percent} of the maximum drift time). For the final event selection, a cut removing S2 events with widths above \SI{1}{\micro\second} is applied, resulting in an event rate of about \SI{750}{\Hz}.

\begin{figure}
\centering
    \includegraphics[width=\linewidth]{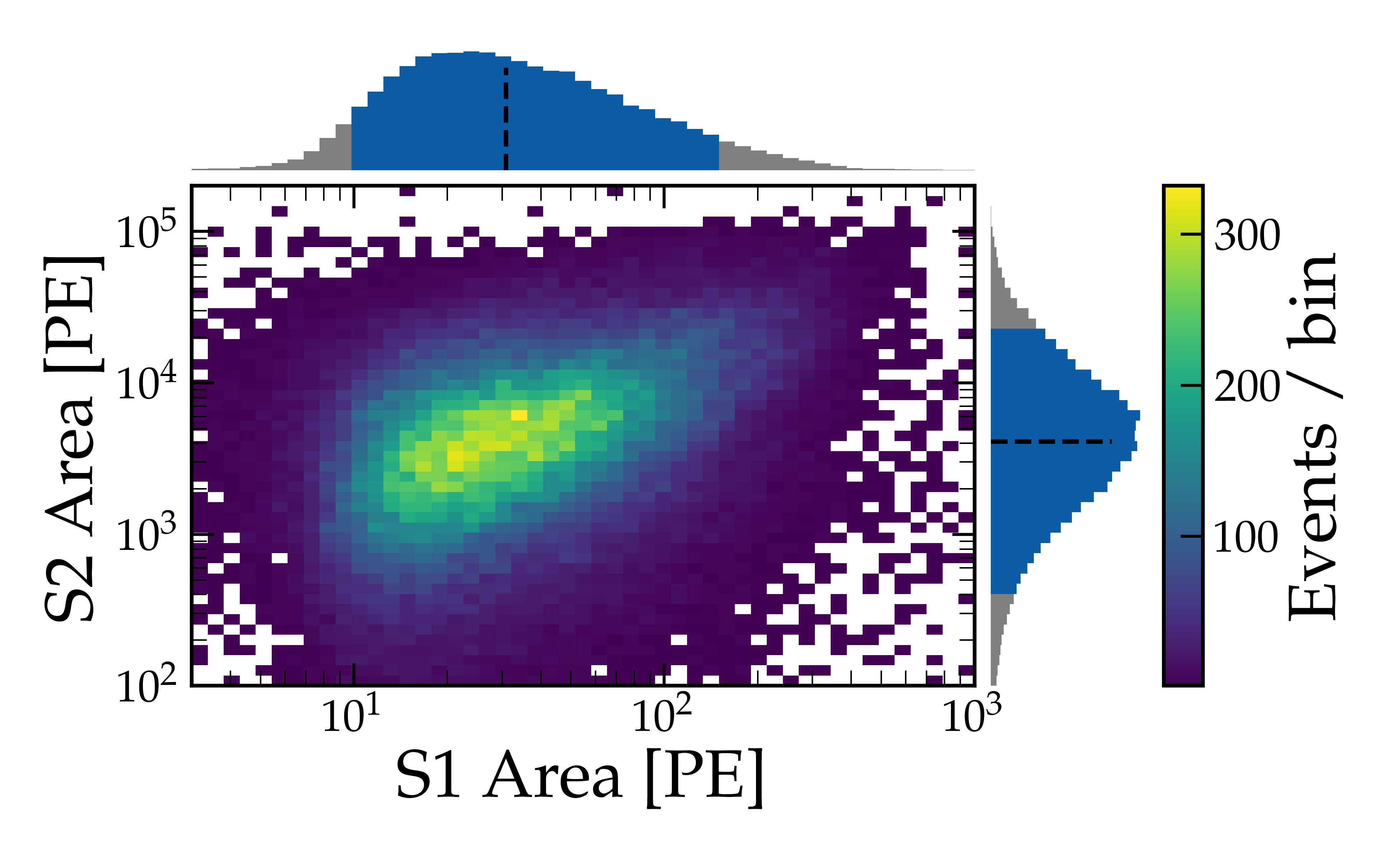}
    \caption{Distributions of the S1 and S2 signal areas of the final dataset after selection cuts. The median S1 (S2) area is \SI{31}{PE} (\SI{4051}{PE}) with the central \SI{90}{\percent} located between [\num{10}, \num{165}]\,\si{PE} for S1s,  and [\num{393}, \num{24137}]\,\si{PE} for S2s). Interestingly, the background spectrum in combination with the very position-dependent detector response leads to peak-like distributions.}
    \label{fig:s1_s2_areas}
\end{figure}

The S1 and S2 signal area distributions of the final event sample are shown in Fig.~\ref{fig:s1_s2_areas}. The median S1 and S2 areas of \SI{31}{PE} and \SI{4051}{PE}, respectively, indicate a relatively low light collection efficiency for prompt and secondary scintillation, consistent with Fig.~\ref{fig:muon_tagger_pos}. For comparison, gamma interactions with $\mathcal{O}$(\SI{1}{MeV}) recoil energies, which make up a large portion of the observed background events, lead to S1 (S2) areas around $10^4$\,PE ($6 \times 10^5$\,PE) in XENONnT~\cite{XENON:2024wpa}. 
Assuming negligible light reflection, the light collection efficiency is primarily determined by the solid angle subtended by the photosensors. Given the shallow TPC geometry, no significant difference is expected between S1 and S2 signals.
The geometric collection efficiency for light emitted at the LXe surface decreases from about \qty{6.5}{\percent} at the TPC center to about \qty{0.2}{\percent} at the outer edge.

%%%%%%%%%%%%%%%%%%%%%%%%%%%%%%%%%%%%%%
\subsubsection{Event waveforms and signal patterns}

\begin{figure}[b!]
    \centering
    \includegraphics[width=\linewidth]{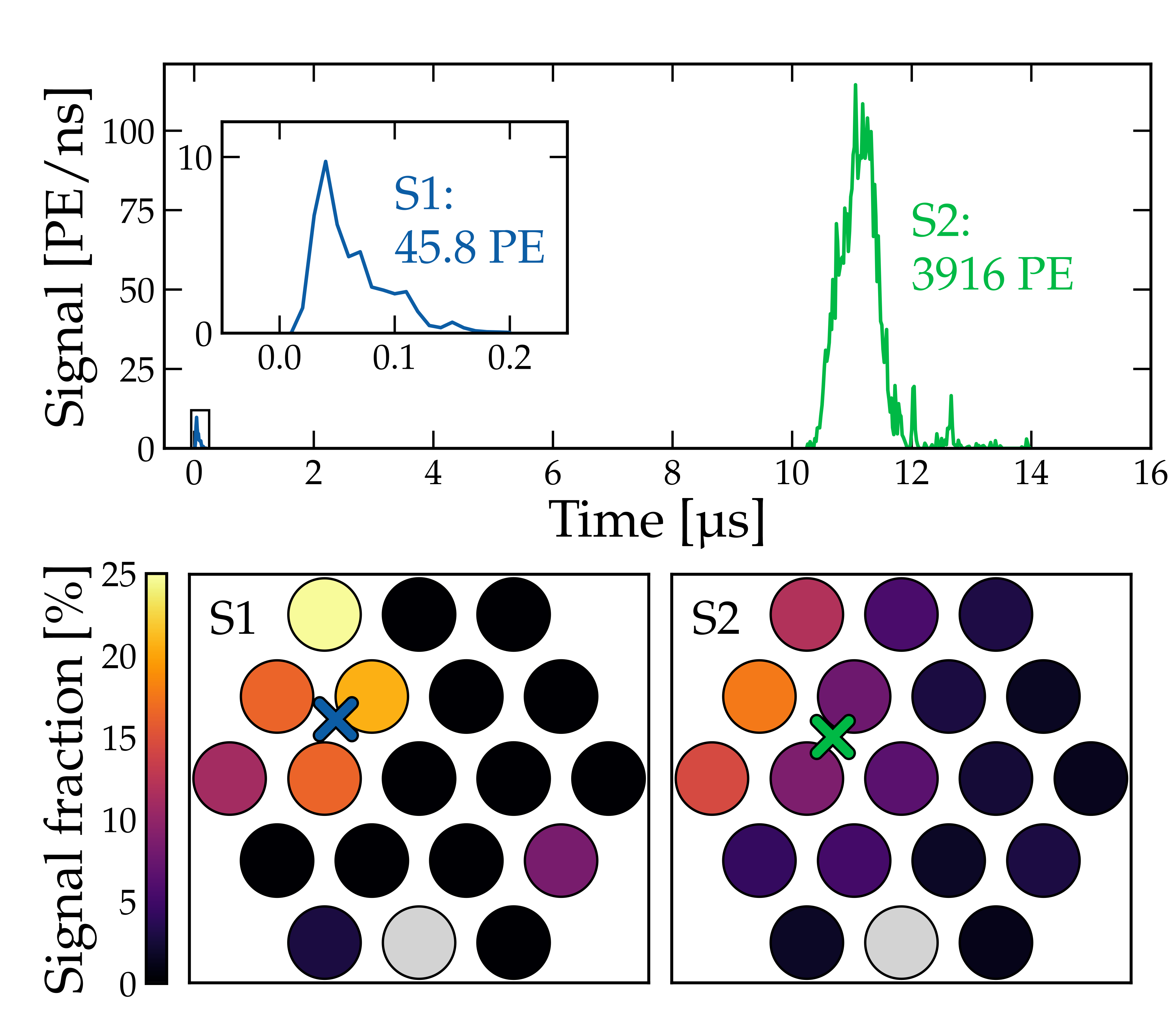}
    \caption{Example of a typical waveform detected by the TPC. The bottom panel shows the distribution of the S1 and S2 signals across the PMTs. The almost identical location of the distribution's centroids (crosses) indicate a common position in the $x$-$y$-plane, as expected for a properly reconstructed event.}
    \label{fig:event_display}
\end{figure}

A waveform of a typical event consisting of an S1 and the much larger S2 signal is shown in Fig.~\ref{fig:event_display}. The drift time, defined as the time difference between the centers of both peaks indicates a rather central vertical $z$-position of the event. The hit patterns in the bottom panel show the recorded signal fraction per PMT channel for both peaks. The centroids, defined as the signal-weighted average of the PMT coordinates and indicated by the colored crosses, show a very good match between the two peaks indicating their similar position in the horizontal $x$-$y$-plane of the TPC. 
A quantitative reconstruction of the horizontal position is not attempted because of the limited area coverage of the small PMT array. 

\begin{figure}
    \centering
    \includegraphics[width=1\linewidth]{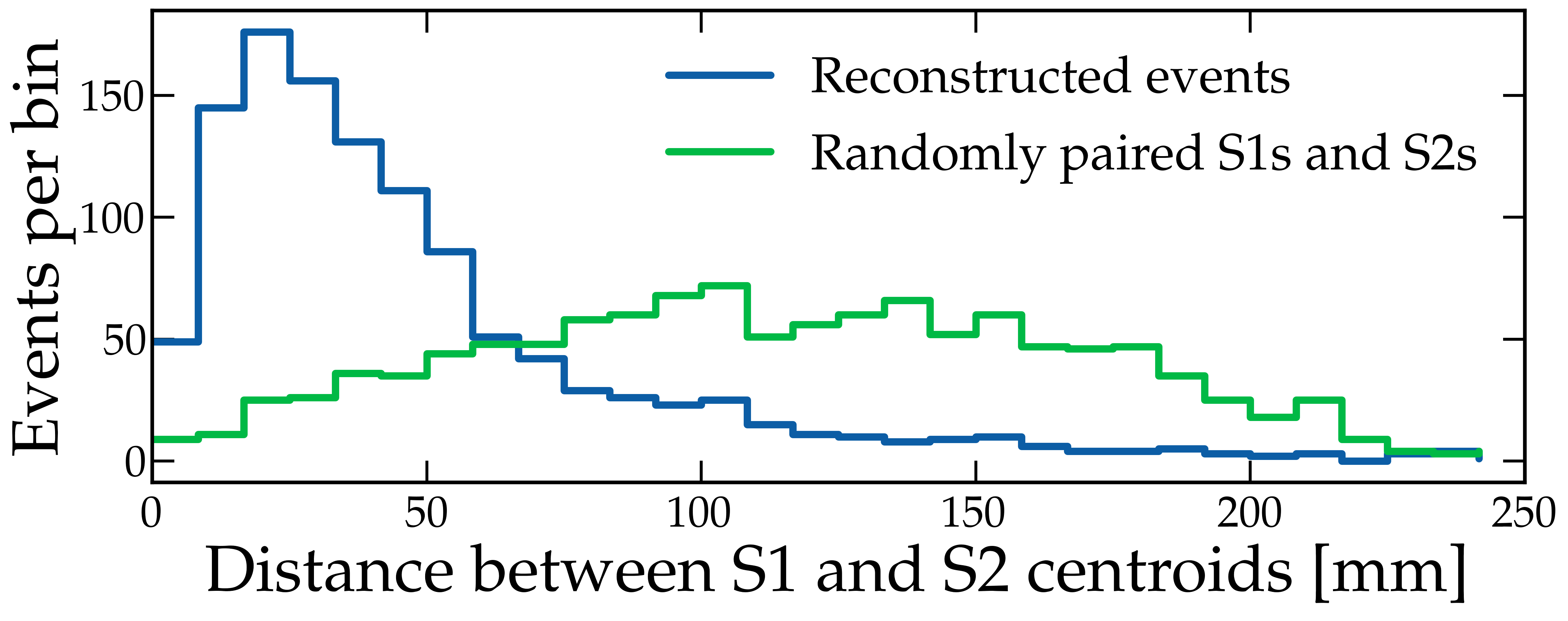}
    \caption{
    The distance between the reconstructed S1 and S2 centroid positions for reconstructed events (blue) exhibits significantly smaller separations than that obtained from randomly paired S1 and S2 signals (green). Only events containing exactly one S1 and one S2 are considered. A minimum S1 size of 50 PE is required, and events with saturated channels are excluded to ensure a valid comparison.}
    \label{fig:posDiff}
\end{figure}

In the TPC, S1 light is produced at the interaction point of a particle, whereas the S2 light is produced in the gas phase below the anode, directly above that point. In the shallow TPC operated here, these two points are always very close together, with the majority of the displacement expected in the vertical direction. Therefore, the distribution of light detected by the PMT array is expected to be similar for these two signals if the S1 and S2 signals are correlated. Figure \ref{fig:posDiff} shows that the spatial distance between the centroid positions of the S1 and S2 signals is strongly correlated for events identified as detailed in Sect.~\ref{sec:ident}. Here, only larger signals are considered, although this behavior does extend to lower S1 sizes until the effect is washed out by low photon statistics. Randomly paired events do not show such correlation, which further validates that the event matching between S1 and S2 peaks works well. 

%%%%%%%%%%%%%%%%%%%%%%%%%%%%%%%%%%%%%%%%%%%%
\subsubsection{Electron lifetime and xenon purity} 
\label{sec:purity}

\begin{figure}
    \centering
    \includegraphics[width=\linewidth]{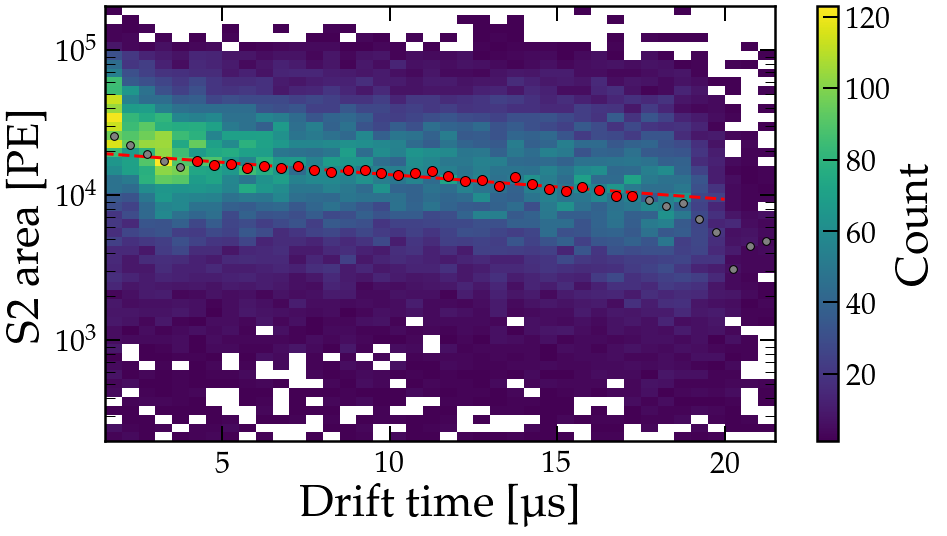}
    \caption{S2 area vs.\ drift time for a dataset with good Xe purity. The red dots indicate the median S2 area of all signals per drift time bin which are described by an exponential decay function (red line). Low drift times are excluded from the fit (grey dots) since the region is affected by the higher extraction field above the gate electrode. The fit yields an electron lifetime of $\tau_e = (25.6 \pm 1.5)\,\SI{}{\micro\second}$.}
    \label{fig:eliftime_fit}
\end{figure}

Liberated electrons from particle interactions may be captured by electronegative impurities (e.g., O$_2$ molecules) along their drift path, leading to a depth-dependent exponential attenuation of S2 signal sizes. The attenuation constant, the so-called electron lifetime $\tau_e$, is a good indicator of the xenon purity in the TPC. 
The event selection described in the previous section is applied to all datasets acquired at different times, and $\tau_e$ is individually determined by fitting an exponential function to the median S2 area per drift time bin, see Fig.~\ref{fig:eliftime_fit}. The resulting lifetimes $\tau_e$ remain consistent for datasets taken within a few hours, indicating the feasibility of this method, which is typically applied to mono-energetic calibration samples rather than background events. During a period of ten days where xenon was constantly purified via a hot-metal getter at \SI{8.5}{SLPM} gas purification flow, an increase in $\tau_e$ from about \SI{10}{\micro\second} to \SI{25}{\micro\second} was observed. This demonstrates that the current purification capabilities of PANCAKE as summarized in Sect.~\ref{sec::setup} are sufficient to achieve a decent charge collection efficiency inside the shallow TPC with a maximal drift time around \SI{20}{\micro\second} (see Fig.~\ref{fig:drifttime_hist}). The electron lifetime measurements were also compared to the H$_2$O concentration the xenon gas measured by the trace gas analyzer a clear qualitative anti-correlation between the two quantities.

%%%%%%%%%%%%%%%%%%%%%%%%%%%%%%%%%%%%%%%%%%%%%%%%%%%%%%
\subsubsection{Electron drift velocity} 

The performance of the PANCAKE TPC can be benchmarked by measuring the electron drift velocity in liquid xenon, which depends on the TPC drift field.  The drift time between gate and cathode electrodes is determined by identifying the corresponding features in the reconstructed drift-time distributions: the cathode position in the time coordinate $t_{\mathrm{cath}}$ is obtained from the sharp cutoff of the drift-time spectrum (see Fig.~\ref{fig:drifttime_hist}), while the gate position $t_{\mathrm{gate}}$ is extracted from the transition in the ratio of S2 to S1 signals as a function of drift time, where the field strength rapidly changes from the drift to the extraction field.  
Both quantities are determined using simple Fermi-like functions, following the method described in Ref.~\cite{xebra}.  The uncertainties include the width of the transition region as well as the statistical fit errors.

The drift velocity is calculated as
\begin{equation} \nonumber
    v_{\mathrm{drift}}
    = \frac{z_{\mathrm{cath}} - z_{\mathrm{gate}}}{t_{\mathrm{cath}} - t_{\mathrm{gate}}} \, ,
\end{equation}
where the $z$-positions are known from detector design. The corresponding drift-field strength is obtained from electrostatic finite element simulations using the median field between gate and cathode within the central detector volume.

\begin{figure}[t]
    \centering
    \includegraphics[width=\linewidth]{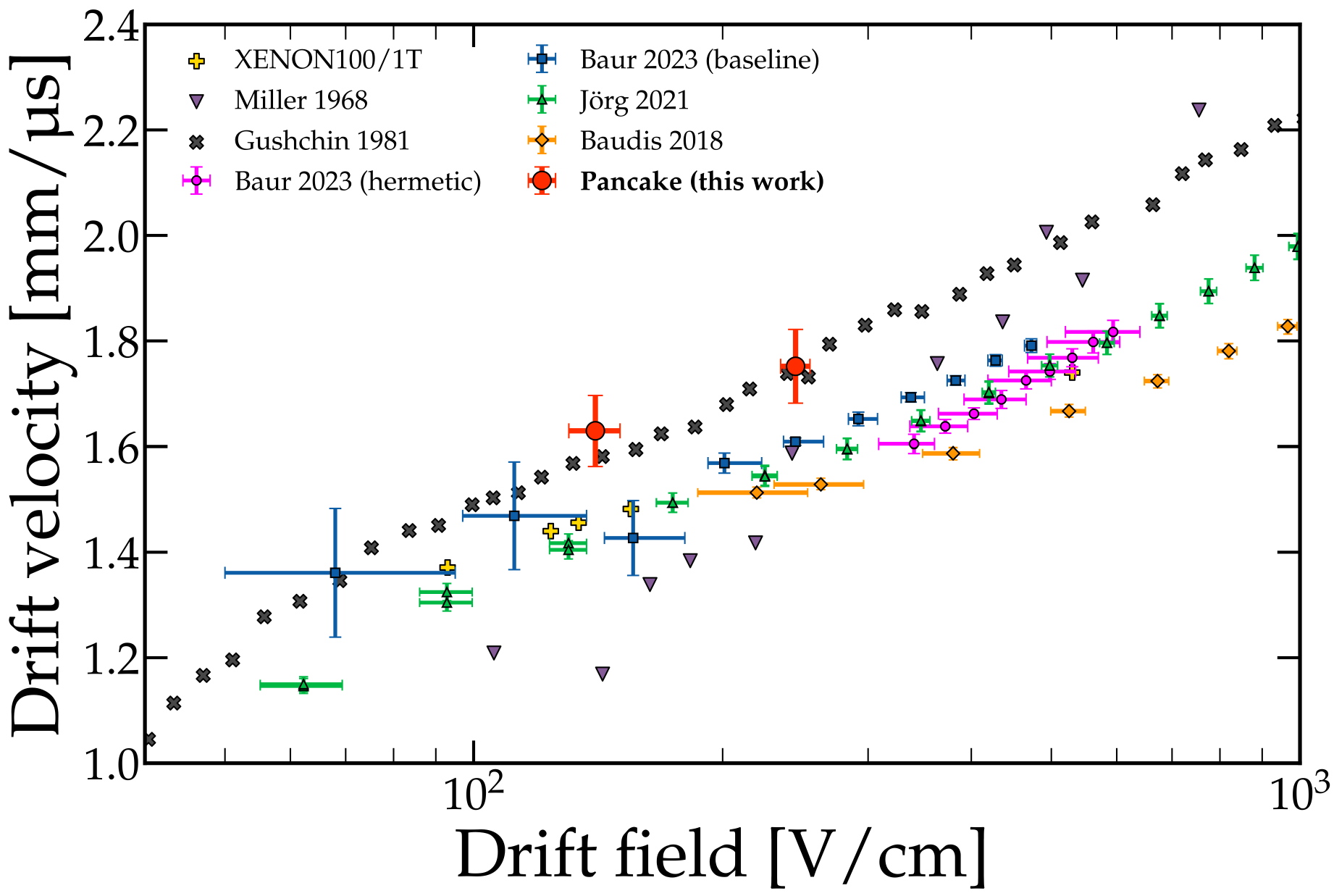}
    \caption{Field-dependent drift velocities measured in PANCAKE (red).  
    Considering the spread typically attributed to differing LXe temperatures, the results agree with previous measurements~\cite{xebra,xenon100,xenon1t,xurich,hexe,miller,gushchin}.}
    \label{fig:v_drift}
\end{figure}

The resulting drift velocities for the two tested PANCAKE TPC fields are shown in Fig.~\ref{fig:v_drift}.  
Within uncertainties, the PANCAKE measurements follow the established field dependence observed in previous experiments. The closest agreement is found with the data of Gushchin \textit{et al.}~\cite{gushchin}, taken at a similar xenon temperature as our measurement ($T=-106^\circ$C), while measurements at higher temperatures yield slightly lower velocities, consistent with the known temperature dependence of the electron mobility in LXe~\cite{xurich}.

%%%%%%%%%%%%%%%%%%%%%%%%%%%%%%%%%%%%%%%%%%%%%%%%%%%%%%%%%%
%%%%%%%%%%%%%%%%%%%%%%%%%%%%%%%%%%%%%%%%%%%%%%%%%%%%%%%%%%
%%%%%%%%%%%%%%%%%%%%%%%%%%%%%%%%%%%%%%%%%%%%%%%%%%%%%%%%%%
\section{Conclusion and Outlook} 
\label{sec::conclusion}

With this work, we have demonstrated that a large, 100-kg-scale shallow LXe TPC with a diameter of about 1.5\,m can be successfully operated for an extended period in an above-ground environment in the PANCAKE platform. The TPC was built from electrodes used to upgrade the XENONnT dark matter detector~\cite{Elykov:2025nri}. 
Despite the absence of an external shield, minimal LXe self-shielding, and no special selection of radiopure construction materials, proper events could be reconstructed from the data. 
The validity of the event reconstruction was verified by studying the spatial correlation of the centroid positions of the signals distributed across the small PMT array. 
The data also enabled measurement of the electron lifetime, which reached values exceeding the maximum drift time of the TPC. Finally, the data were used to measure the field-dependent electron drift velocity.

For the future liquid xenon observatory XLZD, test platforms such as PANCAKE are of crucial importance for developing and testing large-scale components in cryogenic liquid or gaseous xenon. 
While investigating the effects of mechanical stress on large components resulting from cooling by about 120\,K to LXe temperature is an important task in itself, PANCAKE also enables testing of high-voltage systems, such as TPC electrodes.
To this end, the platform was equipped with PMTs, installed in the gas phase above the TPC, and cryogenic cameras. While the cameras are able to identify and locate high-voltage induced sparks, the array of 19~PMTs is sensitive to much fainter light emissions down to the single photon level. However, in this first test, the PMT-based light detection system was not optimized. In particular, no light reflectors were installed, resulting in a rather low light collection efficiency in the center of the TPC, which further decreased towards the edge. \kr calibration events, with a low-energy line at 9.4\,keV, could be identified in the TPC center in light-only-mode. This corresponds to about 15\,keV when field quenching in a TPC drift field of 200\,V/cm is taken into account. 

The detection threshold will be reduced in future runs by adding light reflectors made from PTFE to increase the light collection efficiency. This should then enable measurement of the electron lifetime also with a \kr source. In addition, injections of $^{222}$Rn and $^{220}$Rn into the Xe gas are planned to explore other specific calibration patterns, such as $^{214}$Bi-$^{214}$Po coincidences and high-energy $\alpha$-lines.

% main text ends here
%%%%%%%%%%%%%%%%%%%%%%%%%%

\begin{acknowledgements}
We would like to thank our technical expert Claudio Schmidt, our mechanical engineer Santiago Ochoa and the mechanical workshop of the Institute of Physics in Freiburg, in particular Ralf Schlegel and Khalid Zarouali with his motivated team of apprentices. We also acknowledge the help of the PlanB Electrodes team in XENONnT, Laura Baudis (UZH) and Manfred Lindner (MPIK) for lending us R11410-21 PMTs, Ricardo Peres (UZH) for the initial design of the PMT array, Christian Weinheimer (Münster) for providing us the HALO+ instrument and his team members Robert Braun and Lutz Althüser for their support. We also acknowledge the local help of several colleagues in context of XENON activities, in particular, Frederic Girard (LPNHE) as well as Peter Györgi, Alexander Deisting, Johannes Merz (Mainz), Sebastian Vetter, Klaus Eitel, Alexey Elykov and Hiu-Sze Wu (KIT). 
This work was supported in part by the German Federal Ministry of Research, Technology and Space through the ErUM-Pro grant 05A23VF1, by the German Research Foundation (DFG) through INST 39/1095-1 FUGG and the Research Training Group RTG2044, and the Structure and Innovation Fund of the state of Baden-Württemberg (SI-BW).
\end{acknowledgements}

%%%%%%%%%%%%%%%%%%%%%%%%%%%%%%%%%%%%%%%%%%%%%%%%%%%%%%%%%%%%%%%%%%%%%%
%% References
%%%%%%%%%%%%%%%%%%%%%%%%%%%%%%%%%%%%%%%%%%%%%%%%%%%%%%%%%%%%%%%%%%%%%%


\begin{thebibliography}{99}

\bibitem{Billard:2021uyg}
J.~Billard et al., 
%``Direct detection of dark matter\textemdash{}APPEC committee report*,''
Rept.\ Prog.\ Phys.\ \textbf{85}, 056201 (2022), arXiv:2104.07634, 
doi:https://doi.org/10.1088/1361-6633/ac5754

\bibitem{XENON:2024wpa}
E.~Aprile et al. (XENON Collaboration),
%``The XENONnT dark matter experiment,''
Eur.\ Phys.\ J.\ C \textbf{84}, 784 (2024), arXiv:2402.10446, doi:https://doi.org/10.1140/epjc/s10052-024-12982-5

\bibitem{LZ:2019sgr}
D.~S.~Akerib et al. (LZ Collaboration),
%``The LUX-ZEPLIN (LZ) Experiment,''
Nucl.\ Instrum.\ Meth.\ A \textbf{953}, 163047 (2020), arXiv:1910.09124, doi:https://doi.org/10.1016/j.nima.2019.163047

\bibitem{PandaX:2024qfu}
Z.~Bo et al. (PandaX Collaboration),
%``Dark Matter Search Results from 1.54\,\,Tonne\textperiodcentered{}Year Exposure of PandaX-4T,''
Phys.\ Rev.\ Lett.\ \textbf{134}, 011805 (2025), arXiv:2408.00664,
doi:https://doi.org/10.1103/PhysRevLett.134.011805

\bibitem{Billard:2013qya}
J.~Billard, L.~Strigari and E.~Figueroa-Feliciano,
%``Implication of neutrino backgrounds on the reach of next generation dark matter direct detection experiments,''
Phys.\ Rev.\ D \textbf{89}, 023524 (2014), arXiv:1307.5458, doi:https://doi.org/10.1103/PhysRevD.89.023524

\bibitem{OHare:2021utq}
C.A.J.~O'Hare,
%``New Definition of the Neutrino Floor for Direct Dark Matter Searches,''
Phys. Rev. Lett. \textbf{127}, 251802 (2021), arXiv:2109.03116, doi:https://doi.org/10.1103/PhysRevLett.127.251802

\bibitem{Schumann:2015cpa}
%Dark matter sensitivity of multi-ton liquid xenon detectors
M.~Schumann et al.,
JCAP \textbf{10}, 016 (2015), arXiv:1506.08309, doi:https://doi.org/10.1088/1475-7516/2015/10/016

\bibitem{PANDA-X:2024dlo}
A.~Abdukerim et al. (PandaX Collaboration),
%``PandaX-xT\textemdash{}A deep underground multi-ten-tonne liquid xenon observatory,''
Sci.\ China Phys.\ Mech.\ Astron. \textbf{68}, 221011 (2025), arXiv:2402.03596, doi:https://doi.org/10.1007/s11433-024-2539-y

\bibitem{XLZD:2024nsu}
J.~Aalbers et al. (XLZD Collaboration), 
%The XLZD Design Book: Towards the Next-Generation Liquid Xenon Observatory for Dark Matter and Neutrino Physics
Eur.\ Phys.\ J.\ C 85, 1192 (2025), arXiv:2410.17137, doi:https://doi.org/10.1140/epjc/s10052-025-14810-w

\bibitem{Aalbers:2022dzr}
%A Next-Generation Liquid Xenon Observatory for Dark Matter and Neutrino Physics
J. Aalbers et al. (DARWIN, LZ, XENON Collaborations at al.), 
J.\ Phys.\ G \textbf{50}, 013001 (2023), arXiv:2203.02309, doi:https://doi.org/10.1088/1361-6471/ac841a

\bibitem{XLZD:2024pdv}
J.~Aalbers et al. (XLZD Collaboration), 
% Neutrinoless Double Beta Decay Sensitivity of the XLZD Rare Event Observatory
J.\ Phys.\ G 52, 045102 (2025), arXiv:2410.19016, doi:https://doi.org/10.1088/1361-6471/adb900

\bibitem{DARWIN:2020bnc}
J.~Aalbers et al. (DARWIN Collaboration),
%``Solar neutrino detection sensitivity in DARWIN via electron scattering,''
Eur.\ Phys.\ J. C \textbf{80}, 1133 (2020), arXiv:2006.03114, doi:https://doi.org/10.1140/epjc/s10052-020-08602-7

\bibitem{Brown:2023vgf}
A.~Brown et al., 
%PANCAKE: a large-diameter cryogenic test platform with a flat floor for next generation multi-tonne liquid xenon detectors
JINST \textbf{19}, P05018 (2024), arXiv:2312.14785, doi:https://doi.org/10.1088/1748-0221/19/05/P05018

\bibitem{Kastens:2009pa}
L.W.~Kastens et al., % S.~B.~Cahn, A.~Manzur and D.~N.~McKinsey,
%``Calibration of a Liquid Xenon Detector with Kr-83m,''
Phys. Rev. C \textbf{80} (2009) 045809, arXiv:0905.1766, doi:https://doi.org/10.1103/PhysRevC.80.045809

\bibitem{Hannen:2011mr}
V.~Hannen et al.,
% Limits on the release of Rb isotopes from a zeolite based 83mKr calibration source for the XENON project
JINST \textbf{6}, P10013 (2011), arXiv:1109.4270, doi:https://doi.org/10.1088/1748-0221/6/10/P10013

\bibitem{Zappa:2016zsn}
P.~Zappa et al., % L.~B{\"u}tikofer, D.~Coderre, B.~Kaminsky, M.~Schumann and M.~von Sivers,
%``A versatile and light-weight slow control system for small-scale applications,''
JINST \textbf{11} (2016) T09003, doi:https://doi.org/10.1088/1748-0221/11/09/T09003, arXiv:1607.08189.

\bibitem{Elykov:2025nri}
A.~Elykov et al., 
%``Development {\&} Characterization of Electrodes for large-scale Xenon Time Projection Chambers,''
arXiv:2511.16408.

\bibitem{XENON:2015ara}
E.~Aprile et al. (XENON Collaboration), %"{Lowering the radioactivity of the photomultiplier tubes for the XENON1T dark matter experiment}",
Eur.\ Phys.\ J.\ C \textbf{75}, 546 (2015), arXiv:1503.07698, doi:https://doi.org/10.1140/epjc/s10052-015-3657-5


\bibitem{XENON:2022vye}
E.~Aprile et al. (XENON Collaboration), 
%The Triggerless Data Acquisition System of the XENONnT Experiment
JINST \textbf{18}, P07054 (2023), arXiv:2212.11032, doi:https://doi.org/10.1088/1748-0221/18/07/P07054

\bibitem{Prall:2010}
M.~Prall et al.,
%A Contact-Less 2-Dimensional Laser Sensor for 3-Dimensional Wire Position and Tension Measurements
IEEE Trans.\ Nucl.\ Sci.\ \textbf{57}, 787 (2010), doi:10.1109/TNS.2010.2042612.

\bibitem{pmt_afterpulse_ref}
L.~Baudis et al., 
%``Performance of the Hamamatsu R11410 Photomultiplier Tube in cryogenic Xenon Environments,''
JINST \textbf{8}, P04026 (2013), arXiv:1303.0226, doi:https://doi.org/10.1088/1748-0221/8/04/P04026

\bibitem{strax} J.~Aalbers \emph{et al.}, AxFoundation/strax: v1.5.4 Zenodo, 2023. 

\bibitem{straxen} J.R.~Angevaare et al., XENONnT/straxen: v2.1.1 Zenodo, 2023.

\bibitem{jenks_natural_breaks}
G.~F.~Jenks,
\emph{The Data Model Concept in Statistical Mapping},
Int.\ Yearb.\ Cartogr.\ \textbf{7} (1967) 186--190.

\bibitem{quenching}
E.~Aprile et al.,
% Observation of Anticorrelation between Scintillation and Ionization for MeV Gamma Rays in Liquid Xenon
Phys.\ Rev.\ B \textbf{76}, 014115 (2007), arXiv:0704.1118, doi:https://doi.org/10.1103/PhysRevB.76.014115

\bibitem{xebra}
D.~Baur et al.,
J.\ Instrum.\ \textbf{18}, T02004 (2023), doi:https://doi.org/10.1088/1748-0221/18/02/T02004.

\bibitem{xenon100}
E.~Aprile et al. (XENON Collaboration),
Astropart.\ Phys.\ \textbf{35}, 573 (2012), arXiv:1107.2155, doi:https://doi.org/10.1016/j.astropartphys.2012.01.003


\bibitem{xenon1t}
E.~Aprile et al. (XENON Collaboration),
Eur.\ Phys.\ J.\ C \textbf{77}, 881 (2017), arXiv:1708.07051, doi:https://doi.org/10.1140/epjc/s10052-017-5326-3


\bibitem{gushchin}
E.M.~Gushchin, A.A.~Kruglov, and I.M.~Obodovskii,
JETP \textbf{55}, 650 (1982).

\bibitem{xurich}
L.~Baudis et al.,
Eur.\ Phys.\ J.\ C \textbf{78}, 351 (2018), doi:https://doi.org/10.1140/epjc/s10052-018-5801-5.

\bibitem{hexe}
F.~Jörg et al.,
Eur.\ Phys.\ J.\ C \textbf{82}, 361 (2022), doi:https://doi.org/10.1140/epjc/s10052-022-10259-3.

\bibitem{miller}
L.S.~Miller,
Phys.\ Rev.\ \textbf{166}, 871 (1968), doi:https://doi.org/10.1103/PhysRev.166.871.


\end{thebibliography}
\end{document}